\documentclass{aa}
\usepackage{graphicx}
\usepackage{natbib}
\def\aap{A\&A}

\def\apj{ApJ}

\def\mnras{MNRAS}
\begin{document}
   \title{Hot Stars Mass-loss studied with Spectro-Polarimetric
  INterferometry (SPIN)}

\titlerunning{Hot Stars Mass-loss studied with SPIN}

   \author{O. Chesneau
          \inst{1}
          \and
          S. Wolf\inst{2}
          \and
          A.~Domiciano~de~Souza \inst{3}
          }

   \offprints{O. Chesneau}

   \institute{Max-Planck-Institut f\"{u}r Astronomie,
   K\"{o}nigstuhl 17, D-69117 Heidelberg, Germany\\
              \email{chesneau@mpia-hd.mpg.de}
         \and
             California Institute of Technology,
         1200 E California Blvd., Mail code 105-24,
         Pasadena, CA 91125, USA\\
             \email{swolf@astro.caltech.edu}
         \and
D\'{e}partement d'Astrophysique de l'Universit\'{e} de
Nice/Sophia-Antipolis, CNRS UMR 6525, France\\
\email{Armando.Domiciano@obs-azur.fr}
             }

   \date{Received; accepted }

   \abstract{
   We present a prospective work undertaken on Spectro-Polarimetric
INterferometry (SPIN). Our theoretical studies suggest that SPIN
is a powerful tool for studying the mass loss from early type
stars where strong Thomson scattering is present. Based on Monte
Carlo simulations, we computed the expected SPIN signal for
numerous hot star spectral types covering a broad range of
geometries and optical depths. The SPIN technique is based on the
detection and comparison of the fringe characteristics (complex
visibility) between two perpendicular directions of polarization.
The most obvious advantage is its ability to determine the
polarization distribution in spherical winds for which no
detection of polarization is achievable by classical techniques.
In particular, we demonstrate that the SPIN technique is very
sensitive to the $\beta$ parameter from the so-called '$\beta$
velocity law' for optically thin winds. Moreover, the location
where the bulk of polarization is generated can be defined
accurately. The required sensitivity for studying main sequence OB
star winds is still very demanding (inferior to 0.5\%), but the
signal expected from denser winds or extended atmospheres is well
within the capabilities of existing interferometers. The
visibility curves obtained in two perpendicular polarizations for
LBVs or WR stars can differ by more than 15\%, and their
corresponding limb-darkened radii obtained by the fit of these
curves by more than 35\%. The signal expected from the extended
circumstellar environment of Be stars and B{[}e{]} appears also to
be easy to detect, relaxing the required instrumental accuracy to
1\%. For these spectral types, the SPIN technique provide a good
tool to extract the highly polarized and spatially confined
envelope contribution from the bright star emission.

It must be pointed out that the astrophysical environments
investigated here offer a large panel of SPIN observing conditions
in terms of geometry and polarization degree. The behavior of the
SPIN observables can be transposed, at least qualitatively, to
other astronomical objects for which important local polarization
is foreseen.
   }

   \maketitle
%

\section{Introduction}
Mass-loss is an intrinsic characteristic of hot stars which eject
a strong wind during their whole short life. The light from the
central star can be strongly polarized by its close circumstellar
environment, essentially by Thomson scattering. The mass ejection
is mainly driven by the pressure of the intense radiation field
mediated by
 resonant scattering.
The young technique of optical interferometry has proven its
efficiency to study the close environment of hot stars such as Be
star disks~\cite{Stee95,Quirrenbach97,Vakili98,Berio99} or
environment of the Luminous Blue Variable (LBV) P
Cyg~\cite{Vakili97}. Without spatial resolution,
spectro-polarimetry represents one of the best suited techniques
to study any departure from spherical symmetry of the mass-loss
(see for instance Taylor et al. 1991 and Wood et al. 1997). The
detection of a jet-like structure in the binary $\beta$ Lyrae with
an interferometer (cf. Harmanec et al. 1996) and a
spectropolarimeter (cf. Hoffman et al. 1998) illustrates the
complementarity of both techniques.

However, the interpretation remains limited by the averaging of
the polarized information over the field of view since any
observation of nearly symmetrical object provides an almost
undetectable signal. For instance, the precision of current photo-
and spectro-polarimetric observations is insufficient to test wind
models with latitudinal dependance of the mass-loss rate in O
supergiants, which predict a continuum polarization of only 0.1
per cent at most (Harries et al. 2002).

Within this context, it appears very attractive to equip a long-
baseline interferometer with a polarimetric mode in order to apply
the so-called Spectro-Polarimetric INterferometry (SPIN)
technique. Such attempts have been performed since the very
beginning of interferometry. The unique Narrabri intensity
interferometer was used with a polarimeter in 1974 to give an
estimate of the polarization-dependent diameter change of $\beta$
Orionis~\cite{Hanbury1974}, but the signal-to-noise ratio (SNR)
limitations were well above the expected signal. The experiment
was repeated in 1981 with the I2T interferometer on $\alpha$
Lyrae~\cite{Vakili81}, and in 1997 with the GI2T on the Be star
$\gamma$ Cassiopeiae~\cite{Perraut97}. These observations showed
that instrumental polarization has to be carefully studied and
controlled~\cite{Perraut97}. The first theoretical studies on SPIN
were performed in the frame of the Narrabri Interferometer
experiment by Sams \& Johnston (1974). Cassinelli \&
Hoffman~\cite{Cassinelli75} investigated the consequences of
Thomson scattering around hot stars on the diameter measurements
in linearly polarized light (with a single baseline). In the outer
regions of the star, the light becomes polarized perpendicular in
a direction parallel to the limb of the star. Integrated over the
apparent disk this polarization cancels out. In contrast to this,
an interferometer in polarization mode can detect a signal due to
its sensitivity to the polarized flux in a preferred direction.
The star appears smaller in the plane of polarization parallel to
the baseline than in the plane perpendicular to it. An
illustration of this effect can be seen in Fig.~\ref{fig:refstar},
that shows theoretical iso-intensity contours on the disk of the
star for two orientations of polarization. Rousselet-Perraut
(1998) performed a theoretical study of the SPIN observables based
on simple models with spherical and elliptical scattering
environments. He also presents a methodology that
is useful to interpret the results in the paper.\\
We intend to present an updated overview of the SPIN capabilities
using first, state-of-the-art Monte Carlo simulations, and second
a review of the on-going projects in the field of long-baseline
optical interferometry (i.e.\ instrumental facilities, signal
performances to be obtained or expected in the near future). We
develop a set of astrophysical examples in the field of hot star
winds which covers a broad range of wind geometries and intrinsic
parameters such as the wind density structure. In
Sect.~\ref{interfobs}, the observables provided by long-baseline
optical interferometry are presented, together with the ones more
specific for the signal study in polarized light provided by the
SPIN technique. Sect.~\ref{sec:MC3D} gives a brief description of
the Monte Carlo code MC3D used and adapted for this purpose. In
Sect.~\ref{sect:sphe}, we deal with spherical winds and perform
numerical tests for stars with different spectral types showing
significant winds ranging from A supergiants to Wolf-Rayet stars.
In Sect.~\ref{sec:2D}, we examine 2D geometries, ranging from
anisotropic radiative winds to the disks of Be stars. We then
discuss instrumental polarized devices foreseen and needed for
such a technique in Sect.~\ref{sect:ins}. Finally, we present the
conclusions of this work.
\section{SPIN description}
 \label{interfobs}
In this section we describe the formalism applied in this article.
At first, we briefly recall the interferometric observables
extracted from natural light. We restrict ourself to the case of a
single interferometric baseline, i.e. with two telescopes. We
adopt the formalism of Domiciano de Souza et al. (2002) and
reproduce here the equations necessary for an introduction to
natural light interferometry. In Sect.~\ref{sect:pol} this
formalism is then extended to the case of polarized light.

\subsection{Natural light}
\label{sect:nat} 
\begin{figure}
  \centering
        \includegraphics[height=6.5cm]{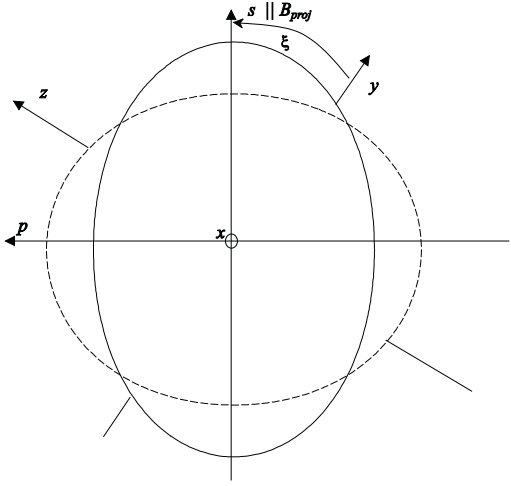}
  \caption{Adopted reference system. The figure represents
  isocontours on the apparent disk of a spherical wind
  as seen with a polarizer parallel (dashed line) and perpendicular (solid one) to the
  sky projected baseline $\vec{B_{\mathrm{proj}}}$. This axis forms
an angle $\xi$ with the sky coordinate system $(x,y,z)$ and
defines a new sky projected coordinate system $(s,p)$ for which
the $s$ direction is parallel to $\vec{B_{\mathrm{proj}}}$. This
$(s, p)$ system is also the frame used in the polarization
analysis.}
  \label{fig:refstar}
\end{figure}

We consider a spherical star defined by its
hydrostatic radius $R_{\rm c}$, located at the center of the
Cartesian coordinate system {\em (x, y, z)} shown in
Fig.~\ref{fig:refstar}. The $y$ axis is defined as the North-South
celestial orientation and the $x$ axis points towards the
observer.

Let us define the sky-projected monochromatic brightness
distribution $I_{\lambda}(y,z)$, hereafter called "natural light
intensity map". Interferometers measure the complex visibility,
which is proportional to the Fourier transform of
$I_{\lambda}(y,z)$. By denoting the Fourier transform of the
intensity map by $\widetilde{I}_{\lambda}(y,z)$ we can write the
complex visibility in natural light as:
\begin{equation}\label{eq:V}
V(f_{y},f_{z},\lambda)=\left| V(f_{y},f_{z},\lambda)\right|
\mathrm{e}^{\mathrm{i}\phi(f_{y},f_{z},\lambda)}=\frac{\widetilde{I}_{\lambda}(f_{y},f_{z})}{
\widetilde{I}_{\lambda }(0,0)},
\end{equation}
where $f_{y}$ and $f_{z}$ are the Fourier spatial frequencies
associated with the coordinates $y$ and $z$. In long-baseline
interferometry the spatial frequencies are given by
$\vec{B_{\mathrm{proj}}}\,\lambda_\mathrm{eff}^{\,\,\textrm{-}1}$,
where $\lambda_\mathrm{eff}$ is the effective wavelength of the
spectral band considered and $\vec{B_{\mathrm{proj}}}$ is a vector
representing the baseline of the interferometer projected onto the
sky. The vector $\vec{B_{\mathrm{proj}}}$ defines the $s$
direction, which forms an angle $\xi $ with the $y$ axis so that:
\begin{equation}\label{eq:Bproj}
\vec{B_{\mathrm{proj}}}=\left( B_{\mathrm{proj}}\cos \xi \right)
\widehat{y}+\left( B_{\mathrm{proj}}\sin \xi \right) \widehat{z}.
\end{equation}
where $\widehat{y}$ and $\widehat{z}$ are unit vectors.

We consider linear cuts along the Fourier plane corresponding to a
given baseline direction $\widehat{s}$. We can define the new
spatial frequency coordinates (u,v) for which
$\vec{B_{\mathrm{proj}}}$ is parallel to the unit vector
$\widehat{u}$. In that case the line integral (or strip intensity)
of $I_{\lambda }(s,p)$ over $p$ for a given $\xi$ can be written
as:
\begin{equation}\label{eq:FTline}
\widetilde{I}_{\lambda,\xi}(u)= \int I_{\lambda,\xi}(s)
\mathrm{e}^{-\mathrm{i}2\pi su} \mathrm{d}s,
\end{equation}
The \textit{ complex visibility} is given by:
\begin{equation}\label{eq:Vline}
V_\xi(u,\lambda)=\left| V_\xi(u,\lambda)\right|
\mathrm{e}^{\mathrm{i}\phi_\xi(u,\lambda)}=\frac{\widetilde{I}_{\lambda,\xi}(u)}{
\widetilde{I}_{\lambda,\xi}(0)}\,.
\end{equation}
By varying the spatial frequency
(baseline length and/or wavelength), we obtain the so-called visibility curve.
Eqs. \ref{eq:FTline} and \ref{eq:Vline} say that the
interferometric information along $\vec{B_{\mathrm{proj}}}$ is
identical to the one-dimensional Fourier transform of the curve
resulting from the integration of the brightness distribution in
the direction perpendicular ($\widehat{p}$) to this baseline.



\subsection{Polarized light}
 \label{sect:pol}

SPIN allows to derive the geometry of the source as detected with
the filtering view of the baseline and the polarization. First, it
must be stressed that contrary to classical polarimetry, the
polarizer direction is {\it not} fixed in the celestial
North-South direction ($y$) but related to the baseline direction
on the sky ($s$ direction). This is due to the fact that the
baseline is fixed to the ground, and not to a moveable mount as in
the case of a monolithic telescopes. Thus it appears natural (and
it is technically straightforward) that the polarization analysis
is using the baseline coordinate system ($s$, $p$) showed in
Fig.~\ref{fig:refstar}. Throughout the entire article, two
particular directions for this polarizer are considered: polarizer
in $s$ direction (parallel to the baseline) and polarizer in $p$
direction (perpendicular to baseline).

Let $I_{\rm lin}$ represent the polarized light contribution of
the intensity map $I_{\lambda}(y,z)$. Our approach is adapted from
the Stokes formalism to our time variable coordinate system of
polarization analysis. We simplify the notation by writing $I_{\rm
nat}=I_{\lambda}(y,z)$. We define
\begin{equation}
I_{\rm lin}=I_{\rm parallel}-I_{\rm perpendicular},
\hspace*{5mm}{\rm and}
\end{equation}
\begin{equation}
I_{\rm s}=I_{\rm nat}-I_{\rm lin}
\end{equation}
\begin{equation}
I_{\rm p}=I_{\rm nat}+I_{\rm lin}.
\end{equation}
When the baseline direction is coincident with the celestial
North-South direction, the polarization analysis system is
coherent with the Stokes formalism: $I_{\rm lin}=I_{\rm Q}$.

The polarized visibility that is measured by a single baseline is
defined by the Fourier transform of the strip intensity of the
intensity maps modulated by the polarization (as defined in
Sect.~\ref{sect:nat}). When we observe with a polarimetric device,
we record (simultaneously or not) three quantities: the visibility
amplitudes curve $|V|$ in natural light ($|V_{\rm nat}|$) and
polarized light ($|V_{\rm s}|$ and $|V_{\rm p}|$). These
visibilities are related to the corresponding intensity maps
$I_{\rm s }$ and $I_{\rm p}$ by relations equivalent to Eqs.
~\ref{eq:FTline} and \ref{eq:Vline}. These visibilities can then
be related to a radius provided that a simple model of the object
light distribution is defined (see Sect.~\ref{sec:numdiam}). In
order to study the polarized signal, we concentrate in this paper
on a few observables chosen for their sensitivity and their
simplicity for interpreting the geometry of the source.

We define the polarized deviation curve $\Delta V_{\rm P}(f)$ as
the difference between the visibility curves in polarized light:
\begin{equation}\label{eq:Vpol}
\Delta V_{\rm P}(f_{y},f_{z},\lambda) = |V_{\rm
p}(f_{y},f_{z},\lambda)| - |V_{\rm s}(f_{y},f_{z},\lambda)|.
\end{equation}
We can also define the degree of
polarized visibility (following the formalism from
Rousselet-Perraut, 1997):
\begin{equation}\label{eq:PV}
P_{\rm V}(f_{y},f_{z},\lambda) = \frac{|V_{\rm
p}(f_{y},f_{z},\lambda)| - |V_{\rm
s}(f_{y},f_{z},\lambda)|}{|V_{\rm nat}(f_{y},f_{z},\lambda)|}.
\end{equation}
The quantities $\Delta V_{\rm P}$ and $P_{\rm V}$ are obtained at
a given time for a given projected baseline. Three regions of the
visibility curves are of common interest in stellar
interferometry: the first lobe, the first minimum and the second
lobe's maximum.

The second lobe's maximum is very sensitive to the limb-darkening
law of the star, and is consequently particularly interesting for
the study of diffuse light. However, precise observations in these
high spatial frequencies require long integration times in order
to compensate the low fringe contrast. Moreover, as far as hot
stars are concerned, the baselines required for studying the
second lobe are generally larger than 200\,m in the NIR band,
except for the few closest stars. This is also true for the first
minimum, commonly used for accurate radius determinations.

Thus, it is more realistic to concentrate on the first lobe. We
define the spatial frequency $f_{\rm max}$ as the frequency where
the SPIN signal $\Delta V_{\rm P}(f_{\rm max})=\Delta V_{\rm max}$
is maximum. The spatial frequency is expressed in units of the
inverse stellar radius, $1/R_*$ and in the figures $R_*=R_{\rm
c}$. As seen in the following sections, $f_{\rm max}$ occurs at
relatively low spatial frequencies, in the $f=(0.2-0.6)$ $1/R_c$
range, which relaxes the spatial resolution needed to perform SPIN
observations drastically.

\section{The Monte Carlo code}
\label{sec:MC3D}
\subsection{Presentation}
The simulation of visibilities and polarimetric observables are
based on radiative transfer simulations performed with the Monte
Carlo radiative transfer code MC3D (Wolf 2003; see also Wolf et
al. 1999, Wolf \& Henning 2000). We assume a spherical, extended
star which radiates isotropically, i.e., the radiation
characteristic at each point of the stellar surface follows the
standard cosine law.  The radiation field of the star is
partitioned into "weighted photons" each of which is characterized
by its wavelength and Stokes parameters. The interaction of the
stellar photons with the surrounding electron envelope is
described by Thomson scattering. Due to (multiple) scattering
events the polarization state of the initially unpolarized photons
is modified. In order to derive spatially resolved images of the
I, Q, and U Stokes vector components of the configuration, photons
leaving the electron envelope are projected onto observing planes
oriented perpendicular to the path of the photons. Since the
optical depth in the electron envelope in some cases becomes $\ll
1$, the enforced scattering concept introduced by Cashwell \&
Everett~(1959) was applied in order to achieve a high
signal-to-noise ratio for the simulated images within a reasonable
computing time. This concept has been used in particular in the
study of the wind of $\zeta$ Puppis presented in
Sec.~\ref{sect:ZP}, and also in Sect.~\ref{sect:epsori} and
Sect.~\ref{sect:deneb}.

\begin{table*}[]
\caption[]{Some relevant parameters for the adopted models of
spherical winds. The targets cover the range of spectral type for
which strong local polarization are expected. Most of the
parameters are adapted from Lamers \& Cassinelli (1999), except
for $\alpha$ Lyre and for WR 40 with parameters from Aufdenberg et
al. 2002 and Herald et al. 2001.}\label{ta:models}
 \vspace{0.1cm}
  \begin{tabular}{lllccccc}
  \hline
 & & & $\zeta$ Puppis & $\epsilon$ Ori & Deneb &
  P Cyg & WR 40\\
\hline
Type  & & & O4If & B0Ia & A2Iae & B1Ia/LBV & WN8\\
    Distance& $D$ &  pc & 430  & 410 & 685 & 1800 & 2260\\
    Stellar radius& $R_c$ & ${\rm R}_\odot$ & 17  &  35 & 172 & 76 & 11\\
    Core angular diameter & $\Theta$&  mas & 0.35 & 0.8 & 2.35 & 0.4 & 0.04\\
    Stellar temperature& $T_*$& K & 42000 & 28000 & 8875 & 19300 & 45000\\
    Mass loss rate & $\dot{M}$& ${\rm M}_\odot\ {\rm yr}^{-1}$ & $6\times10^{-6}$ & $4\times10^{-6}$ &$1\times10^{-6}$ &$1.5\times10^{-5}$ &$3\times10^{-5}$\\
    Terminal velocity & $v_\infty$ & ${\rm km}\ {\rm s}^{-1}$ & 2200 & 1500 &225&210 & 840\\
    Acceleration coefficient & $\beta$ & & 1.0 & 1.5 & 3 & 2.5 & 1\\
    Optical depth & $\tau_{\rm e}$ & & 0.2 & 0.17 &0.03 & 1 & 3.4\\
    \hline
  \end{tabular}

\end{table*}

\subsection{Limits}
\label{sect:lim} In this study, an important limitation is that
MC3D determines the polarization due to multiple photon scattering
by electrons, but does not include the effects of continuous
hydrogen absorption and emission seen in disk-like circumstellar
envelopes for example. Consequently, our modelling of a hydrogen
disk in Sect.~\ref{sec:2D} results in an upper limit for the
expected interferometric signal from an interferometer,
especially for disk studies. This effect is discussed in Sect.~\ref{sect:ins}.\\


\subsection{"Numerical" Diameters}
\label{sec:numdiam}
 For each example treated in this paper, we
perform a fit of the numerical visibility curves obtained with the
code MC3D in order to derive the apparent diameter $\Theta_{\rm
ap}$, which is by definition larger than the diameter $\Theta_{\rm
c}$ defined with the hydrostatic radius $R_{\rm c}$. The
definition of the parameter $\Theta_{\rm ap}$ is not
straightforward but depends on the model used for the fit. If the
centre-to-limb variation (CLV) of intensity could be observed
directly,
 an intensity radius could be defined in terms of the CLV shape and then
 related to a monochromatic optical-depth or filter radius via a model.
In practice, reconstruction of the CLV from
 interferometric data is difficult with presently attainable
 accuracies and the mean number of visibility points observed for
each star. Diameters are usually derived by fitting the visibility
of a well-defined artificial CLV like, e.g. a uniform disk (UD), a
limb-darkened
 disc (LD) or a Gaussian intensity distribution to the observed visibility
 (cf. Jacob \& Scholz 2002). The radius estimation
based on the Uniform Disk (UD) assumption is not sufficient since
we have chosen spectral types for which the diffuse light from
Thomson scattering is important. A better way is to perform a fit
of the visibilities based on the assumption that the emergent
radiation follows a simple cosine limb-darkening law across the
disk of the star (Limb-Darkened disk, LD):
\begin{equation}
I=1-c_{{\rm s/p}}(1-\mu)
\end{equation}
The two parameters of the fitted function are the LD angular
diameter $\Theta_{\rm LD}$ ($\Theta_{\rm LDs}$ or $\Theta_{\rm
LDp}$ in polarized light) and the limb-darkening coefficient $c$
($c_{\rm s}$ or $c_{\rm p}$). These LD radii can be related by the
canonical UD radius by (adapted from Sams \& Johnston 1974):
\begin{equation}
\Theta_{\rm UDs}=\left(\frac{1-\frac{7}{15}c_{\rm
s/p}}{1-c/3}\right)^{(1/2)}\Theta_{\rm LDs}.
\end{equation}

\begin{figure}
   \includegraphics[width=8cm]{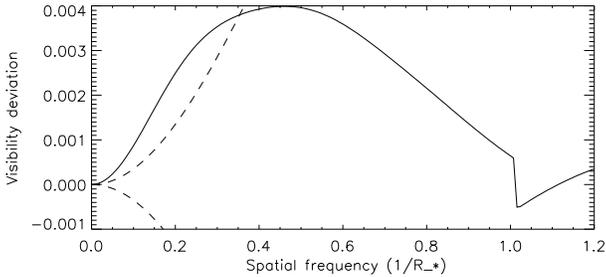}
      \caption[]
  {\label{fig:Deneb}
The expected signal polarization from the Deneb wind model is
relatively weak, the $\Delta V_{\rm P}$ curve is below the 1\%
accuracy illustrated by the dashed curves defined from
Eq.~\ref{eq:sens}. Nevertheless, a signal could be still
detectable owing to the brightness and the angular diameter of
this star.}
\end{figure}

In order to provide a clear view on the instrumental capabilities
offered by contemporary interferometers, we have assumed a
theoretical error curve based on a UD visibility curve estimation.
Inside the first lobe a good analytical approximation for the
visibility uncertainties $\sigma V$ due to the apparent star
radius $R_{\rm ap}$ is given by (Vakili et al. 1997):
\begin{equation}
\label{eq:sens} \frac{{\sigma_{R_{\rm ap}}}} {R_{\rm ap} } =
\frac{{\sigma V }} {{\left| 2 {\mathrm{J}_{2} \left( z \right)}
\right|}}.
\end{equation}
where $\mathrm{J}_{2}$ is the Bessel function of the first kind
and second order. In the following, we assume that the
interferometer is able to constrain $R_{\rm ap}$ with a 1\%
accuracy, and we build the related ${\sigma V }$ curve. This curve
is then overplotted in each $\Delta V_{\rm P}$ diagram (in
Fig.\ref{fig:ZP-standard2}, Fig.\ref{fig:Deneb},
Fig.\ref{fig:PCYG2}, Fig.\ref{fig:WR}, Fig.\ref{fig:ZT},
Fig.\ref{fig:BeC}) as a visual scale of the instrument accuracy
for an illustration purpose.

\section{Spherical winds}
\label{sect:sphe}

In this section, we consider spherical winds, for which the
integrated polarized information averages out to a null value. In
Tab.~\ref{ta:models}, we present a list of typical early type
stars for which the spatial characteristics (i.e.\ mainly the
angular diameter and the brightness) are well suited for
interferometric observations. Moreover, their winds are thick
enough to expect a clear SPIN signal from their diffuse light. It
must be pointed out that the angular diameters reported in this
table are not the expected {\it apparent angular diameters}
$\Theta_{\rm ap}$, but the {\it core diameter} or $\Theta_{\rm c}$
used as input parameter for the radiative transfer simulation. The
expected apparent angular diameter in natural light is increased
by the wind as the star appears larger as a result of the diffused
light from the halo.

For the wind velocity we apply the so-called $\beta$ law:
\begin{equation}
  v(r)=v_\infty\left(1-\frac{r_0}{r}\right)^\beta,
\end{equation}
where
\begin{equation}
  r_0=R_{\rm c}\left[1-\left(\frac{v_0}{v_\infty}\right)^{1/\beta}\right].
\end{equation}
The wind velocity is radial, accelerating to the value $v_\infty$.
The stellar radius $r_0$ is defined as the hydrostatic radius
$R_{\rm c}$ and in the following, $v_0$ is arbitrarily chosen to
be 10
km s$^{-1}$.\\
The mass loss rate is related to the density and the velocity via
the following relation:
\begin{equation}\label{eq:dens}
\dot{M}=4 \pi r^2 \rho (r) v(r).
\end{equation}
The local density is extracted from this relation for a given
$\beta$ law and mass-loss rate.

\subsection{$\zeta$ Puppis}
\label{sect:ZP}

\begin{figure}
 \centering
   \includegraphics[height=6.1cm]{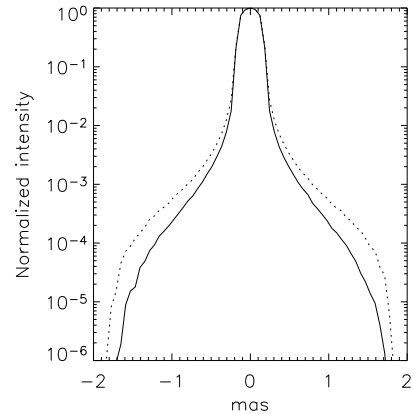}
 \caption{$\zeta$ Puppis flux integrated in the baseline direction (strip intensity map) for polarizations $s$
 (solid line) and $p$ (dotted line). Close to the
photosphere, the wind density strongly decreases following a rapid
acceleration modelled by the $\beta$ law.}
  \label{fig:ZP-standard1}
\end{figure}

In order to illustrate the expected signal from a spherical wind,
we modelled the star $\zeta$ Puppis (HD~66811), an early O4If
supergiant. Davis et al. (1970) have given an estimation of its
apparent angular diameter based on a Uniform Disk fit:
$\Theta_{\rm ap}=0.42\pm0.03$ mas. In the following, we perform a
study of the $\beta$ law parameters and then discuss the $\zeta$
Puppis 'standard' model.

\begin{figure}
 \centering
  \includegraphics[height=7cm]{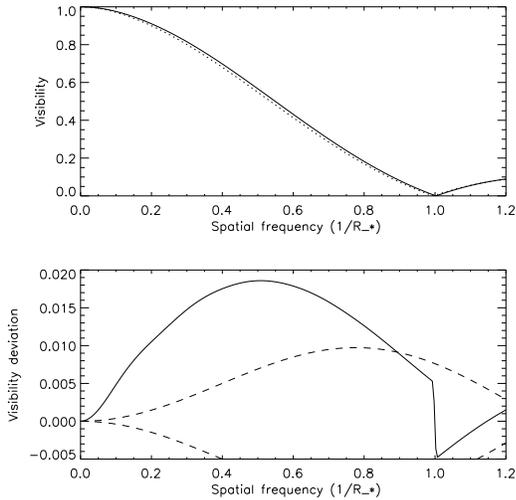}

  \caption{Visibility signal for both polarization, $V_s$ in solid line, $V_p$ in dotted line (top),  and their
difference  $\Delta V_{\rm P}$(bottom) for the adopted parameter
of $\zeta$ Puppis wind. The signal is small, but detectable if the
interferometer sensitivity is such that an accuracy of 1\% on
radii measurements is possible (illustrated by the dashed curves,
from Eq.~\ref{eq:sens}). The discontinuity close to the spatial
frequency $f=$1 appears because the first zero point is located at
lower frequency for the polarization $p$ intensity map, slightly
more extended than the $s$ one. }
  \label{fig:ZP-standard2}
\end{figure}

\subsubsection{'Standard' model}
The parameters of our 'standard model' are described in
Tab.~\ref{ta:models}. Using this model, we find a maximum
visibility deviation $\Delta V_{\rm max}$=0.017 at the spatial
frequency $f_{\rm max}=0.47$. Taking
$\lambda_\mathrm{eff}=$1\,$\mu {\rm m}$ and the angular diameter
of $\zeta$ Puppis ($\Theta_{\rm ap}=0.42$ mas), this corresponds
to a baseline of 300m, but only 180m at $\lambda_\mathrm{eff}=0.6
\mu {\rm m}$. As we can see in Fig.~\ref{fig:ZP-standard2}, such a
SPIN signal is close to the detection limit of an interferometer
able to detect radius deviations of the order of 1\%.

Cassinelli \& Hoffman (1975) also took $\zeta$ Puppis as reference
and gave the first estimate of the SPIN signal for a star with a
wind. They used a two-component density model for which an
extended atmosphere of total thickness $\tau_{\rm e}$=10 is
connected to a flow region with an optical thickness $\tau_{\rm
e}$=0.19, which is close to our standard parameters.
 However, their wind follows a different density law:
\begin{equation}\label{eq:power}
\rho=\rho_0\left(\frac{R_c}{r}\right)^n.
\end{equation}
We have conducted a comparative study with the Cassinelli \&
Hoffman density law ($n=-2$) and the same $\tau_{\rm e}$. Compared
to our standard $\zeta$ Puppis model (with the $\beta$ law), this
new model presents a diffused light multiplied by a factor 2.5.
For an equivalent optical depth, $\Delta V_{\rm max}$ is almost
doubled, and $f_{\rm max}=0.18$ only, to be compared with $f_{\rm
max}=0.47$ in case of our 'standard' model. The polarization is
generated much farther from the star with the power law model and
the optimum baseline to detect the polarized signal taking
$\lambda_\mathrm{eff}=$1~$\mu {\rm m}$ (resp.~$0.6\mu{\rm m}$)
should be reduced to 120\,m (resp.~70\,m).

By performing a least-square fit with a uniform disks to the
simulated visibilities, Cassinelli \& Hoffman expect an angular
diameter ratio between two perpendicular polarization directions
of 7\%, and $\Theta_{\rm ap}$ from a uniform disk fit of the star
emission is 12\% larger than the $\Theta_{\rm c}$. The results
from the 'standard' model are smaller by a ratio of 2\%, and
$\Theta_{\rm ap}$=1.07\,$\Theta_{\rm c}$. The contribution from
the extended atmosphere included in Cassinelli \& Hoffman appears
to strongly favor the polarized signal compared to our Monte Carlo
simulation of the wind component only. Their result has been
confirmed later by Castor, Abbott \& Klein (1975) who estimated
that the scattering halo of $\zeta$ Puppis increases $\Theta_{\rm
c}$ by 13\% with a similar mass-loss rate
($\dot{M}=6.6\times10^{-6}{\rm M}_\odot$/y). Nevertheless a
refined study from Kudritzki et al.(1983) showed that the
atmosphere of $\zeta$ Puppis cannot be considered as extended,
which gives a lower limit $\Theta_{\rm c}=0.38\pm0.03$ mas, and a
diameter increase of less than 10\% in all cases and probably as
low as a few percents, close to the result from the present
'standard' model.

\subsubsection{Mass-loss rate : $\dot{M}$}
\begin{figure*}
 \centering
   \includegraphics[width=14cm]{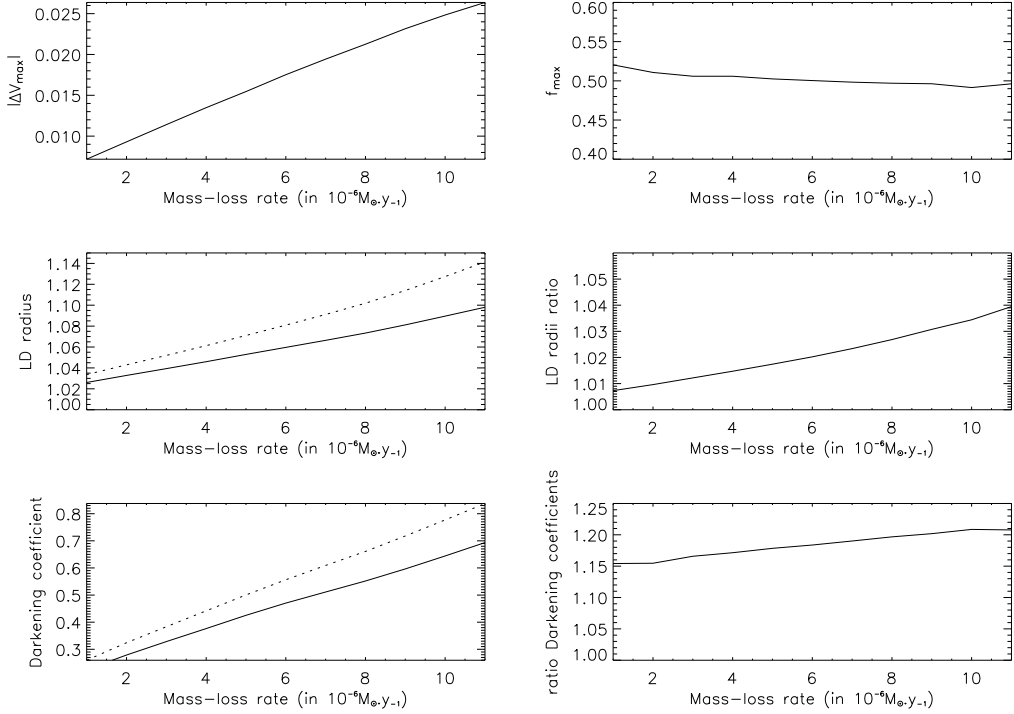}
 \caption{Top: Behavior of the maximum SPIN signal $\Delta V_{\rm max}$ and the corresponding spatial frequency $f_{\rm max}$ with $\dot{M}$.
 Middle and bottom: Result of fits of the polarized visibility curves with a 2 parameters
 limb-darkening law: LD angular diameters $\Theta_{\rm LDs}$ (solid line) and $\Theta_{\rm LDp}$ (dotted line)
 and LD coefficients $u_{\rm p}$
 (solid line) and $u_{\rm s}$ (lower line). In right the ratio  $\Theta_{\rm LDp}$/$\Theta_{\rm LDs}$ and
 $u_{\rm p}$/$u_{\rm s}$ are displayed. As the mass-loss increases, the contrast between $\Theta_{\rm LDp}$ and
 $\Theta_{\rm LDs}$ increases but $u_{\rm p}$/$u_{\rm s}$ evolves more slowly}
  \label{fig:mass}
\end{figure*}

In this section we investigate how the mass-loss rate may affect
the SPIN signal. As expected, we see in Fig.~\ref{fig:mass} that
the SPIN signal $\Delta V_{\rm max}$ follows  the mass-loss rate
increase and the subsequent electron density increase almost
linearly. The wind from $\zeta$ Puppis is optically thin
throughout the entire range of mass-loss rate encompassed. The
spatial frequency $f_{\rm max}$ at which this maximum $\Delta
V_{\rm max}$ can be detected is relatively stable. This means that
the spatial location where the bulk of the polarization is
generated is relatively unaffected by a change of the mass-loss.
However, the limb-darkening coefficients $c_{\rm s/p}$, are very
sensitive to the increase of diffused light since they are
multiplied by a factor larger than 3 as seen in
Fig~\ref{fig:mass}. This means that the relative light
distribution, i.e.\ the balance of diffuse light near and far from
the star evolves with the mass-loss. Slight multiple scattering
effects are visible for high $\dot M$: $\Delta V_{\rm max}$
evolves more slowly with $\dot{M}$.

\subsubsection{Velocity : $v_\infty$}
\begin{figure}
 \centering
   \includegraphics[width=9.cm, height=5.cm]{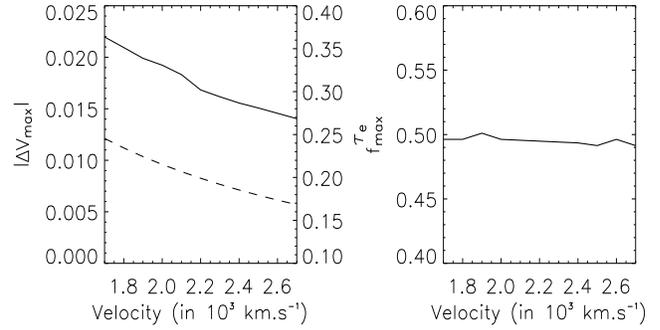}
 \caption{Behavior of the maximum SPIN signal and the corresponding spatial frequency with parameter $v_\infty$.
 The dashed curve represents the course of the optical depth $\tau_{\rm e}$.}
  \label{fig:vel}
\end{figure}

Now $\dot{M}$ and $\beta$ are kept fixed to their 'standard'
values: $\dot{M}$=$6\times10^{-6}$ and $\beta$=1. In
Eq.~\ref{eq:dens}, we see that increasing the terminal velocity
$v_\infty$ will decrease the local electron density and the
scattering. The influence of a change in this parameter on the
local electron density has a similar impact as changing the
mass-loss rate. The behavior of the SPIN signal follows the local
electron density and the electron optical depth reported in
Fig.~\ref{fig:vel}. As mentioned for the mass-loss rate, $f_{\rm
max}$ can be considered as constant. Nevertheless, changing
$v_\infty$ has a greater impact on $\Delta V_{\rm max}$ than a
change of the mass-loss rate: the slope of $\Delta V_{\rm max}$
versus $v_\infty$ in Fig.~\ref{fig:vel} is twice as large as the
slope versus $\dot{M}$ in Fig.~\ref{fig:mass}. The differences
between the mass loss and velocity plots can be understood by the
fact that the optical depth of the wind varies as $\tau_{\rm e}$ =
$\dot{M}$ /$v_\infty$. Thus, $\tau_{\rm e}$ and the visibility
increase proportional to $\dot{M}$ and inversely with $v_\infty$.
Moreover the modification of $v_\infty$ in the $\beta$ law affects
the local density close to the star (where the radiative field is
strong) significantly by changing the value of $r_0$. The
scattering close to the star is increased by a larger amount by a
change of $v_\infty$ than by a change of $\dot{M}$.

\subsubsection{Power-law index : $\beta$}
This parameter defines the density variation from the star to
farther regions. Taking the parameters from the $\zeta$ Puppis
standard model, we have varied $\beta$ between 0.8 to 1.5.

\begin{figure}
 \centering
   \includegraphics[width=9cm, height=5.0cm]{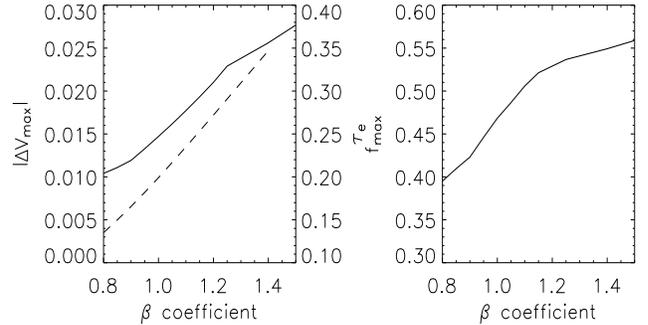}
 \caption{Behavior of the maximum SPIN signal and the corresponding spatial frequency with parameter $\beta$.
 The dashed curve represents the course of the optical depth $\tau_{\rm e}$.}
  \label{fig:beta}
\end{figure}

We see in Fig.~\ref{fig:beta}, that $f_{\rm max}$ and $\Delta
V_{\rm max}$ are strongly sensitive to $\beta$. This demonstrates
that the parameter $f_{\rm max}$ is very sensitive to the main
location of the scattered light: {\it The smaller $f_{\rm max}$,
the flatter the density law, i.e.\ the bulk of the polarized
emission is further away from the star}. This behavior is
particularly interesting since the $\beta$ parameter is usually
difficult to constrain by classical techniques using spectroscopic
data such as spectrum fitting. A change of $\beta$ has often a
small influence on the spectrum itself: the spatial extent of the
line forming regions is affected but the lack of spatial
information from spectroscopy prevents from constraining
efficiently this parameter (see Hillier et al. 1998 for instance).
Harries et al. (2002) pointed out recently the sensitivity of the
polarized line profile morphology of O supergiants to the adopted
velocity field. He suggests that it may be possible to use it as a
diagnostic tool for the wind base kinematics. Such a diagnostic
could also be performed by an interferometer with sufficient
spectral resolution (see discussion in Sect.\ref{sect:ins}).

\subsection{$\epsilon$ Ori}
\label{sect:epsori} The signal expected from $\epsilon$ Ori is
comparable to the one from $\zeta$ Puppis: $v_\infty$ and
$\dot{M}$ are similar, and thus $\tau_{\rm e}$ also. But this star
is more interesting for interferometric observations since the
star is more extended than an O4If star with a similar expected
SPIN signal. With $\lambda_\mathrm{eff}=0.6 \mu {\rm m}$ (resp.
$\lambda_\mathrm{eff}=$1\,$\mu {\rm m}$, the baseline of half
resolution ($f=0.5$) is only 70m (120m). This means that B
supergiants offer a good compromise between polarized signal and
spatial resolution. This is especially true for those which
exhibit the strongest wind manifestations as LBVs
(Sect.~\ref{sect:pcyg}).

\subsection{Deneb}
\label{sect:deneb}
 We performed simulations for Deneb ($\alpha$~Cygni)
based on the exhaustive work of Aufdenberg et al.~\cite{Auf02}.
This A supergiant represents the "cool" detection limit of Thomson
scattering in the wind. The $\beta$ law parameters are estimated
from a fit of the numerical electron density law from Aufdenberg
et al. \cite{Auf02} using Eq.~\ref{eq:dens}. The model parameters
are displayed in Tab.~\ref{ta:models}. The optical depth due to
electron scattering is about 7 times lower than for $\zeta$
Puppis, and the expected ratio between two perpendicular
directions of polarization does not exceed 0.5\%. Nevertheless, we
expect the results from our Monte Carlo simulation to be a lower
limit for the expected SPIN signal produced in the extended
atmosphere of this star (Sect.~\ref{sect:lim}), and the radius
ratio between perpendicular polarizations could be of the order of
1\%. The UD angular diameter $\Theta_{\rm ap}$ from Aufdenberg et
al.~\cite{Auf02} of 2.4\,mas is slightly larger than the core
diameter expected from a non-extended atmosphere. However, it must
be pointed out that Deneb's H$\alpha$ profile exhibits a lack of
the broad emission wing seen in the
spectra of other supergiants, which are normally attributed to electron scattering.\\
In natural light we expect that the apparent diameter of Deneb
remains identical, whatever the baseline direction on the sky may
be. The quasi-sphericity of the envelope is well established based
on the absence of integrated polarization and detectable variation
of radius with baseline for interferometric measurements
(Aufdenberg et al.~2002).

\subsection{P Cyg}
\label{sect:pcyg} LBVs are instable blue supergiants which cross a
short-lived stage of instabilities with large mass loss rate in
the HR diagram. The slow, dense wind from LBVs and in particular
P~Cygni is at least optically thick due to Thomson scattering
($\tau_{\rm e}=1$ for P~Cygni), and multiple scattering occurs.
The observed radius in natural light is very different from the
hydrostatic one, defined by the basic P~Cygni parameters.

\subsubsection{Smooth wind}
\begin{figure}
 \centering
   \includegraphics[height=7.cm]{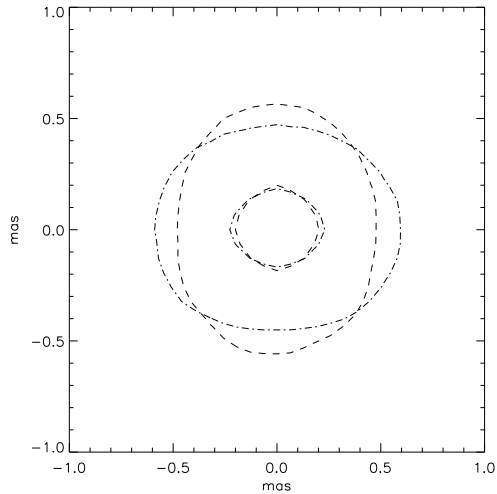}
   \caption[]
  {\label{fig:PCYG1}
Square root of polarized intensity maps contours for two
perpendicular polarizers for the P~Cygni smooth wind model $p$
direction in dashed line and $s$ direction in dashed-dotted line).
The inner contour delimits the level 0.8$I_{\rm max}^2$, where
$I_{\rm max}$ is the maximum of the intensity $I$. It is close to
the core angular diameter of 0.4 mas determined from P~Cygni
parameters defined in Tab~\ref{ta:models}. The outer contour
delimits the level 0.3$I_{\rm max}^2$.}
 \end{figure}

\begin{figure}
 \centering
   \includegraphics[height=8cm]{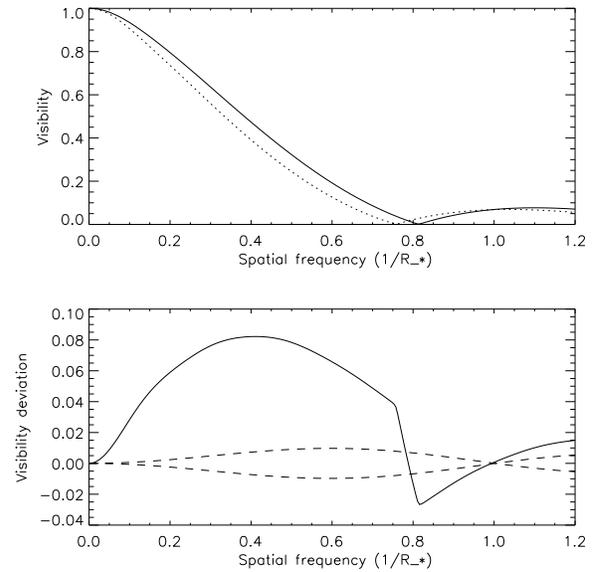}
   \caption[]
  {\label{fig:PCYG2}
Visibility $V_{s}$ (solid line, up) and $V_{p}$ (dotted line, up )
and the SPIN signal $\Delta V_{\rm P}$ (solid line, down) for the
P~Cygni smooth wind model. The SPIN signal is large and well above
the 1\% accuracy curves (dashed lines, down). $\Delta V_{\rm
max}$=0.087 at $f_{\rm max}=0.40$, which corresponds to a relative
signal $P_V$ larger than 12\%. }
 \end{figure}

In Fig.~\ref{fig:PCYG1} the isocontours of the polarized intensity
maps $I_{\rm s}$ and $I_{\rm p}$ are shown. They strongly depart
from spherical symmetry, and this effect increases with the
distance since the local polarization is larger in this case. The
diffuse light is important and the angular diameter of the star in
natural light represents 135\% of the core angular diameter
presented in Tab.~\ref{ta:models}: $\Theta_{\rm ap}=0.55$\,mas
according to the prediction of the detailed P~Cygni model of
Najarro (2001). The SPIN signal is also large as shown in
Fig~\ref{fig:PCYG2}. The optimum baseline to detect the signal is
110\,m (resp.~185\,m) with $\lambda_\mathrm{eff}=0.6~\mu{\rm m}$
(resp.~$1\,\mu{\rm m}$). But the signal is still strong at shorter
baselines: $\Delta V_{P}$=0.05 at a normalized spatial frequency
$f$=0.15 (6\% visibility difference), which corresponds to a 40\,m
baseline for $\lambda_\mathrm{eff}=0.6 \mu {\rm m}$ and 70\,m for
$\lambda_\mathrm{eff}=1 \mu {\rm m}$. The ratio between radii in
perpendicular polarization directions reaches 1.24. The signal
amplitude is well within the accuracy of optical interferometers
which compensates the fact that the star is not well resolved.

Such a strong SPIN signal is also expected for well-known LBVs
such as AG Car, HR Car or HD316285. The last star has spatial
parameters very close to those of P~Cygni ones, i.e. an equivalent
estimated distance, radius or K magnitude, but differs by its huge
mass-loss rate of the order of $2\times10^{-4}{\rm M_{\odot}
yr^{-1}}$, more than 10 times larger than that of P~Cygni (Hillier
et al. 1998). Its wind is very thick ($\tau_{\rm e} \simeq 7$),
and the ratio between radii in perpendicular polarized lights
reaches 1.37. Due to the high level of multiple scattering, the
increase of mass-loss rate does not increase the SPIN signal
compared to the case of P~Cygni ($\Delta V_{\rm max}$=0.08). But
the bulk of polarized light is generated far from the star and the
weight of these extended polarized regions compared to the
essentially unpolarized and heavily damped central star increases
dramatically. The spatial frequency needed for resolving the
polarized halo is consequently much lower compared to the one
needed to resolve the central star, and the SPIN signal peaks at
$f_{\rm max}=0.18$. The apparent angular diameter in natural light
based on the LD fit reaches $3.0\times \Theta_{\rm c}=1.2$~mas
(2.5 for the UD angular diameter). It must be noted that the
visibility curve of HD 316285 is not adequately described by a
Uniform or Limb-Darkened disk visibility fit. For these extreme
spectral types fits based on the assumption of a gaussian
distribution of light have to be preferred.

\subsubsection{Clumps}
The integrated polarization of P~Cygni exhibits a strong
variability ($\sim$0.4\%) on time-scales of days to weeks (Taylor
et al. 1991), with no favored position angle which implies a
quiescent state close to sphericity. This behaviour has been
related to strong and localized eruptions and from an in-depth
study of the spectopolarimetric variability. Nordsieck et
al.~(2001) have strongly restricted the 'polarized' clump
parameter space: position $r<2 R_c$, radius=0.1$R_c$, density
contrast $\rho/\rho_0$=20. These clumps could be related to those
detected further out (about 0.5 arcsec) with Adaptive Optics by
Chesneau et al. 2000.

Previous interferometric studies have demonstrated that the
modulus of the visibility (as used in the paper) is not very
sensitive to small scale asymmetries of the object (Vakili et al.
1997). However, the photocenter of the emission in natural and
polarized light is no longer centered on the star and consequently
affects the phase of the fringes. The expected signal should be
faint but a differential study between two polarizations or
between line and continuum should increase the accuracy. Such a
study for photometry and polarimetry, similar to the one carried
out by Rodrigues \& Magalh\~{a}es~(2000), is not in the scope of
this paper and deserves more extensive investigation.

\subsection{WR 40}
The Wolf-Rayet (WR) evolutionary stage is characterized by a
strong mass-loss rate and a very dense and optically thick wind.
These stars also exhibit a much smaller hydrostatic radius, and a
faster terminal velocity compared to the LBV stage.

Most of the brightest WR stars are located in distances larger
than 1\,kpc and their core radii are generally smaller than
4\,${\rm R}_\odot$ which implies core angular diameters
$\Theta_{\rm c}$ lower than 0.04\,mas. Nevertheless, the expected
apparent diameter can be much larger because the wind is optically
thick far from the star. Among WR subtypes, the WN8 star offers a
good compromise between luminosity and core radius extent, which
reach 10-15 $R_\odot$, and they can therefore be more easily
detected with long-baseline interferometers.

\begin{figure}
 \centering
   \includegraphics[height=8.cm]{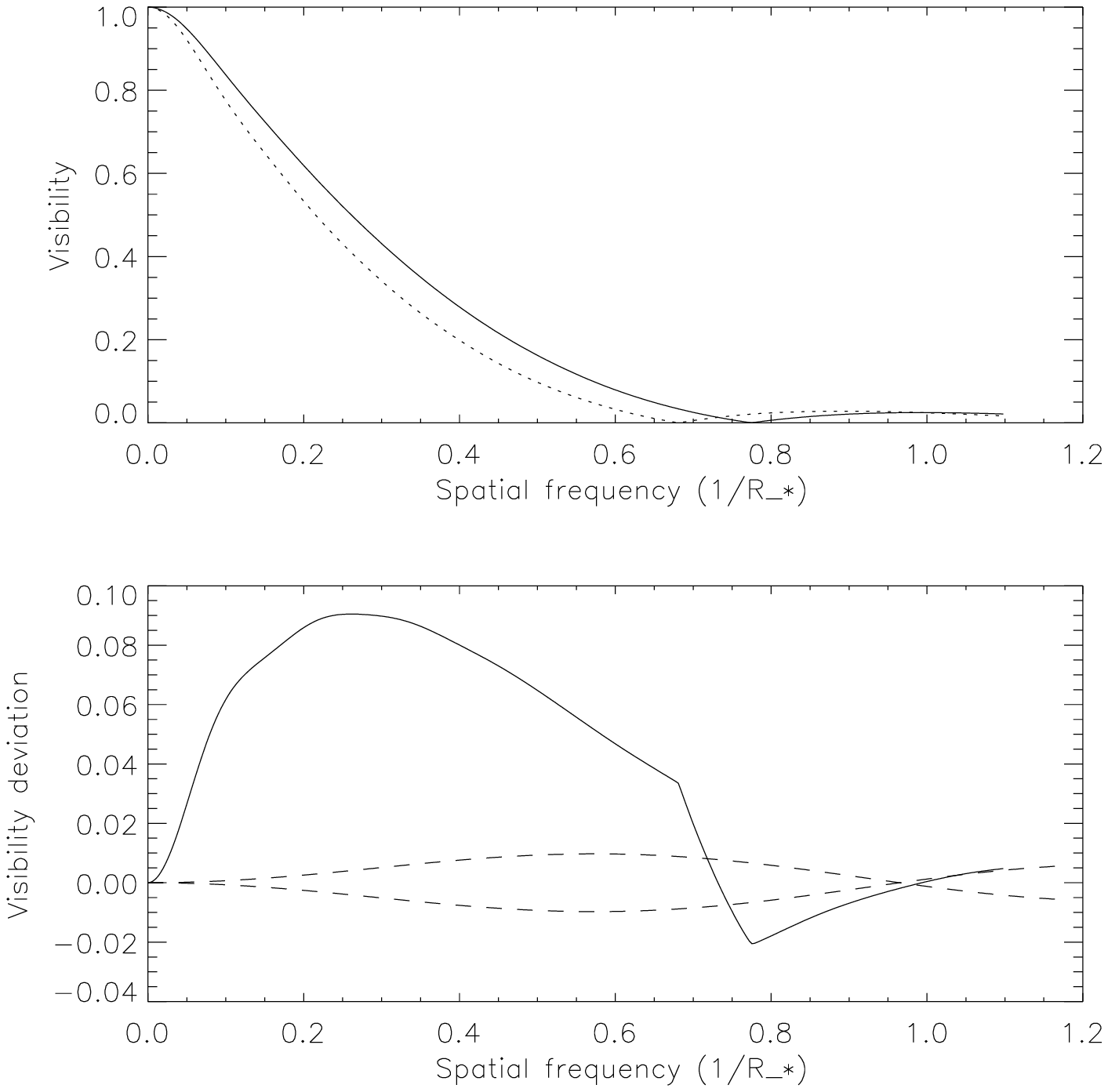}
   \caption[]
  {\label{fig:WR}
Visibility $V_{s}$ (solid line, up) and $V_{p}$ (dotted line, up )
and the SPIN signal $\Delta V_{\rm P}$ (solid line, down) for the
WR~40 model. $\Delta V_{\rm P}$ is well above the curves of 1\%
accuracy (dashed lines, down). The second lobe of the visibility
function almost disappears by apodisation for the WR model because
the central star is no longer detectable within the envelope.}
 \end{figure}

The parameters of WR40 are extracted from the dedicated study of
Herald, Hillier \& Schulte-Ladbeck (2001), and the SPIN signal
presented in Fig~\ref{fig:WR}. As for HD316285 $\Delta V_{\rm
max}$ saturates to the P~Cygni value, but the shape of $\Delta
V_{\rm P}$ is characteristic of a strongly optically thick wind
(see also Cassinelli \& Hoffman 1975): $f_{\rm max}$=0.26, and the
visibility curves can no longer be fitted by UD or LD disk
visibility laws.

The WR wind parameters are well constrained with current
line-blanketed non-LTE models atmospheres. Nevertheless, the
constrains on the $\beta$ parameter are still weak. Herald et
al.~(2001) investigated $\beta$ over a range of 0.5-2 and they did
not detected any significant impact on their spectra. By varying
$\beta$, we notice that the SPIN signal is also no longer
sensitive to this parameter. The optically thick zone is so large
that most of the wind acceleration is embedded in it and the
polarization comes essentially from regions close to the terminal
velocity. The study of the $\beta$ law can only be conducted for
optically thin winds.


\section{Disk geometries}
\label{sec:2D}

In this section, we investigate objects for which the close
environment can no longer be considered as spherically symmetric
and presents a 2D structure created by a colatitude dependance of
the mass-loss rate. The generated structure can be an extended
compressed equatorial region like for B[e] stars or even a disk
for Be stars for instance.

\begin{figure*}
   \includegraphics[height=5.5cm]{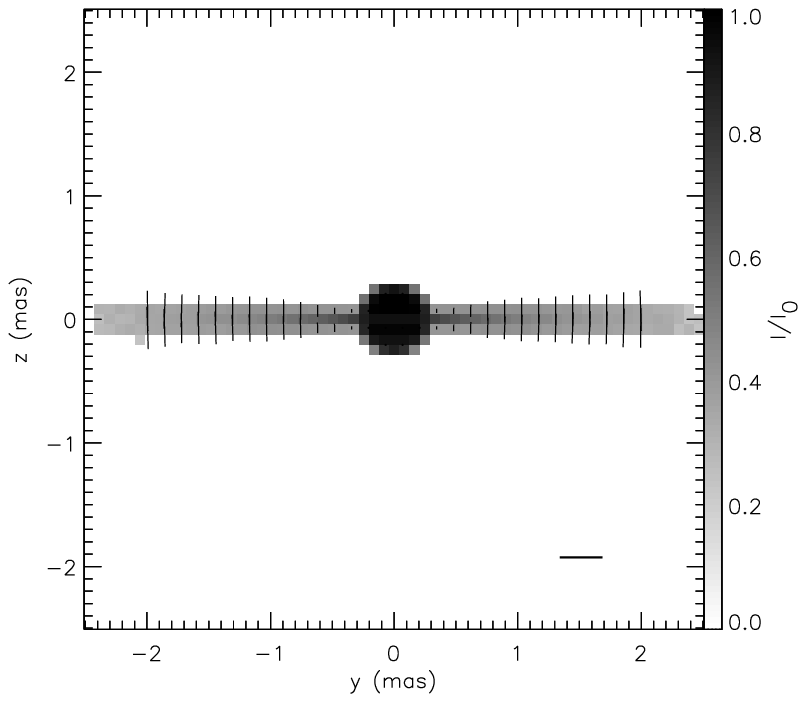}
   \includegraphics[height=5.5cm]{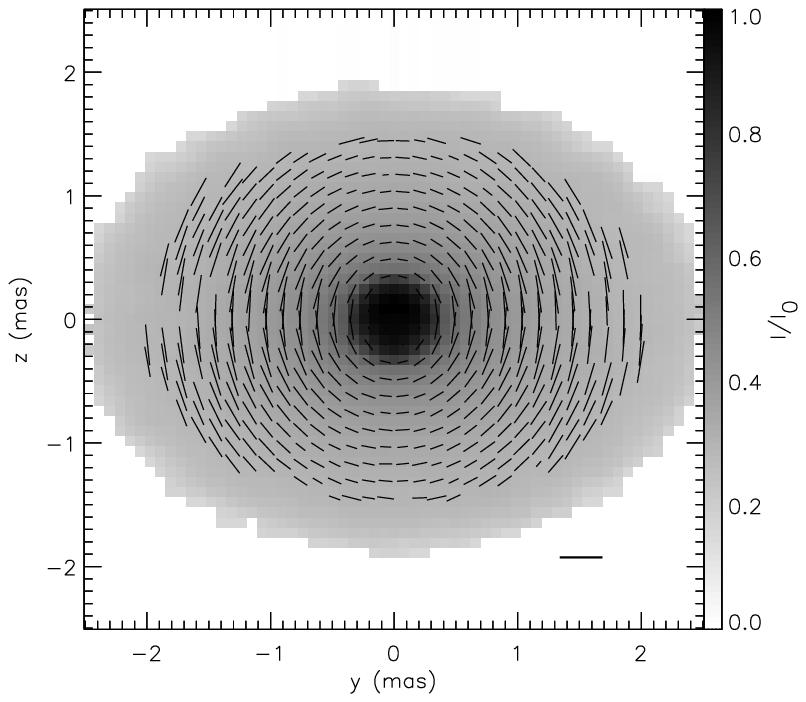}
      \includegraphics[height=5.5cm]{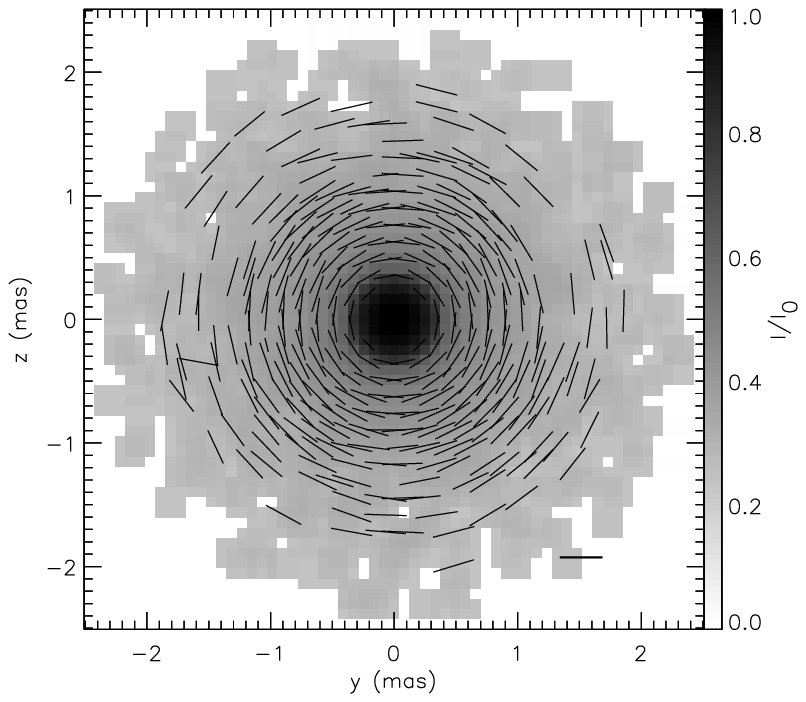}\\
        \includegraphics[height=5.3cm]{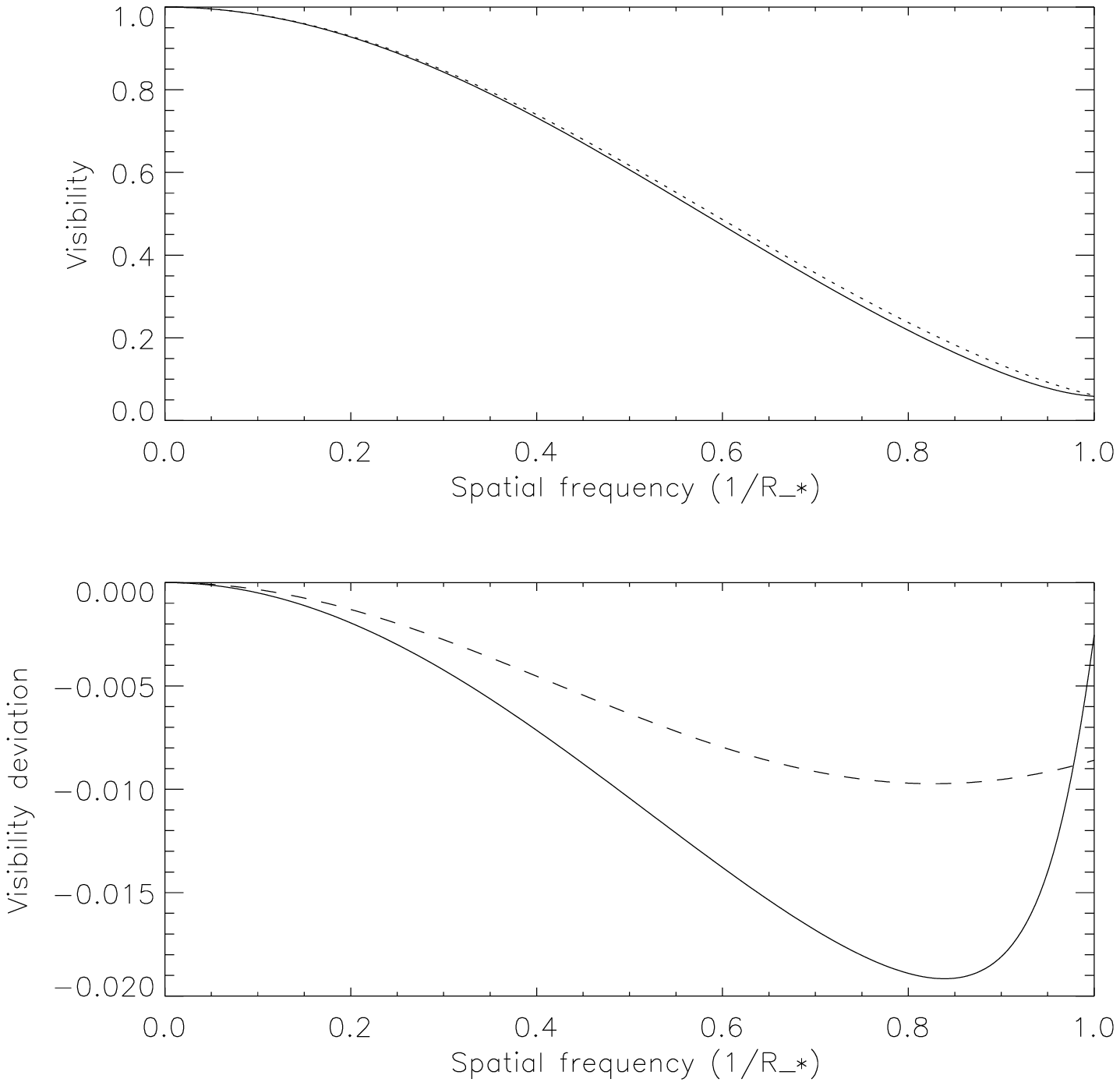}
 \includegraphics[height=5.3cm]{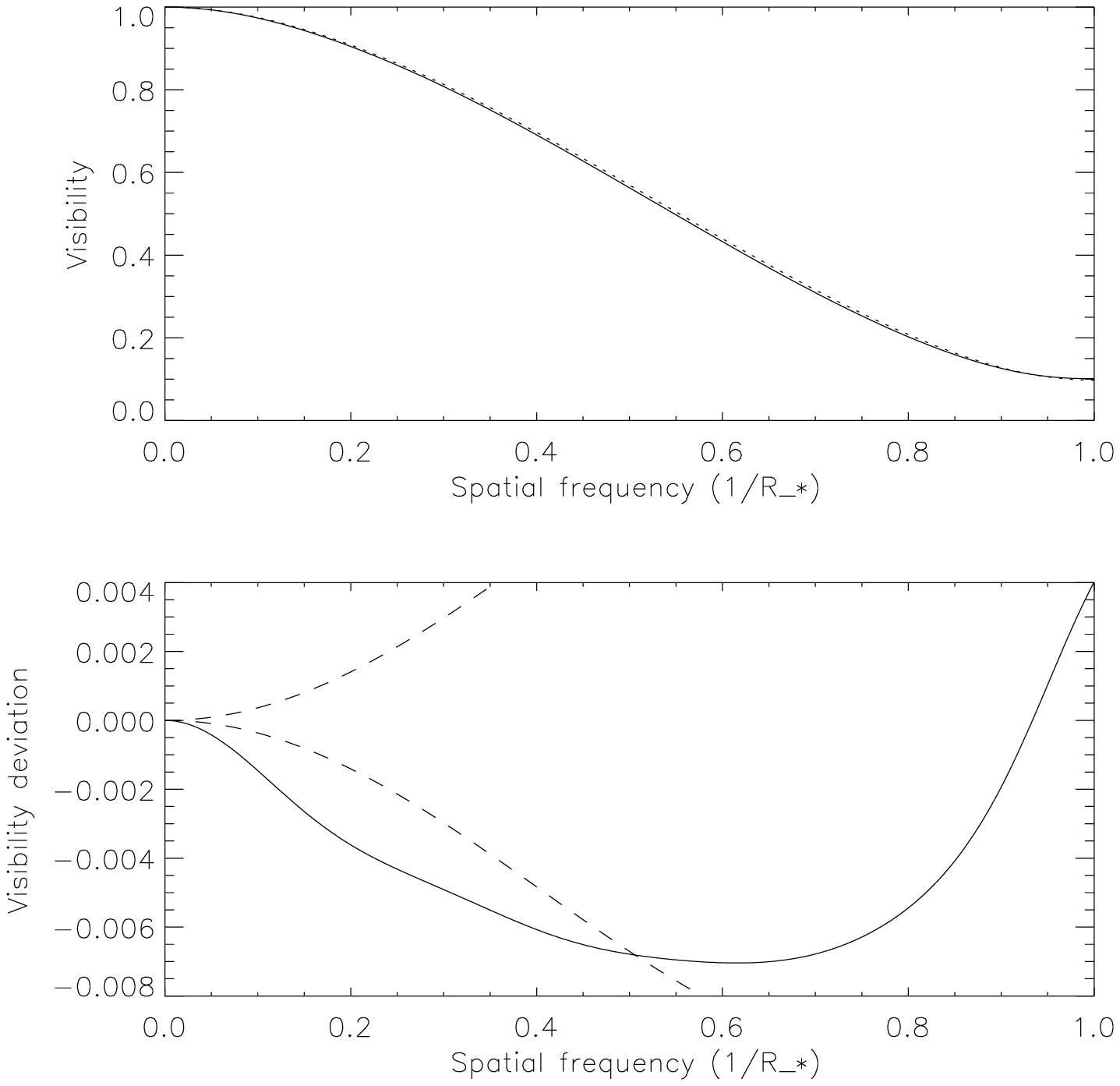}
   \includegraphics[height=5.3cm]{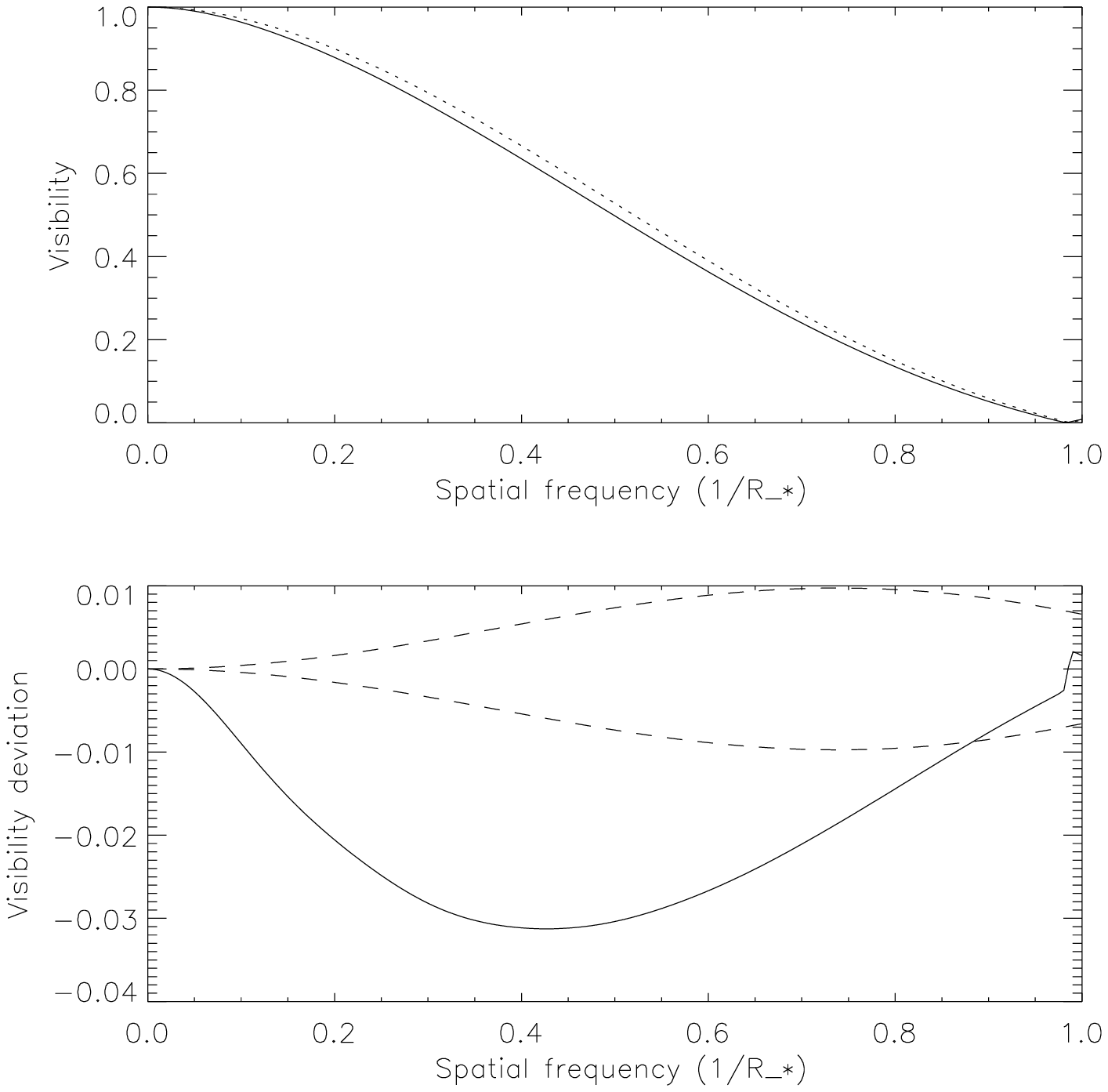}\\
    \includegraphics[height=5.3cm]{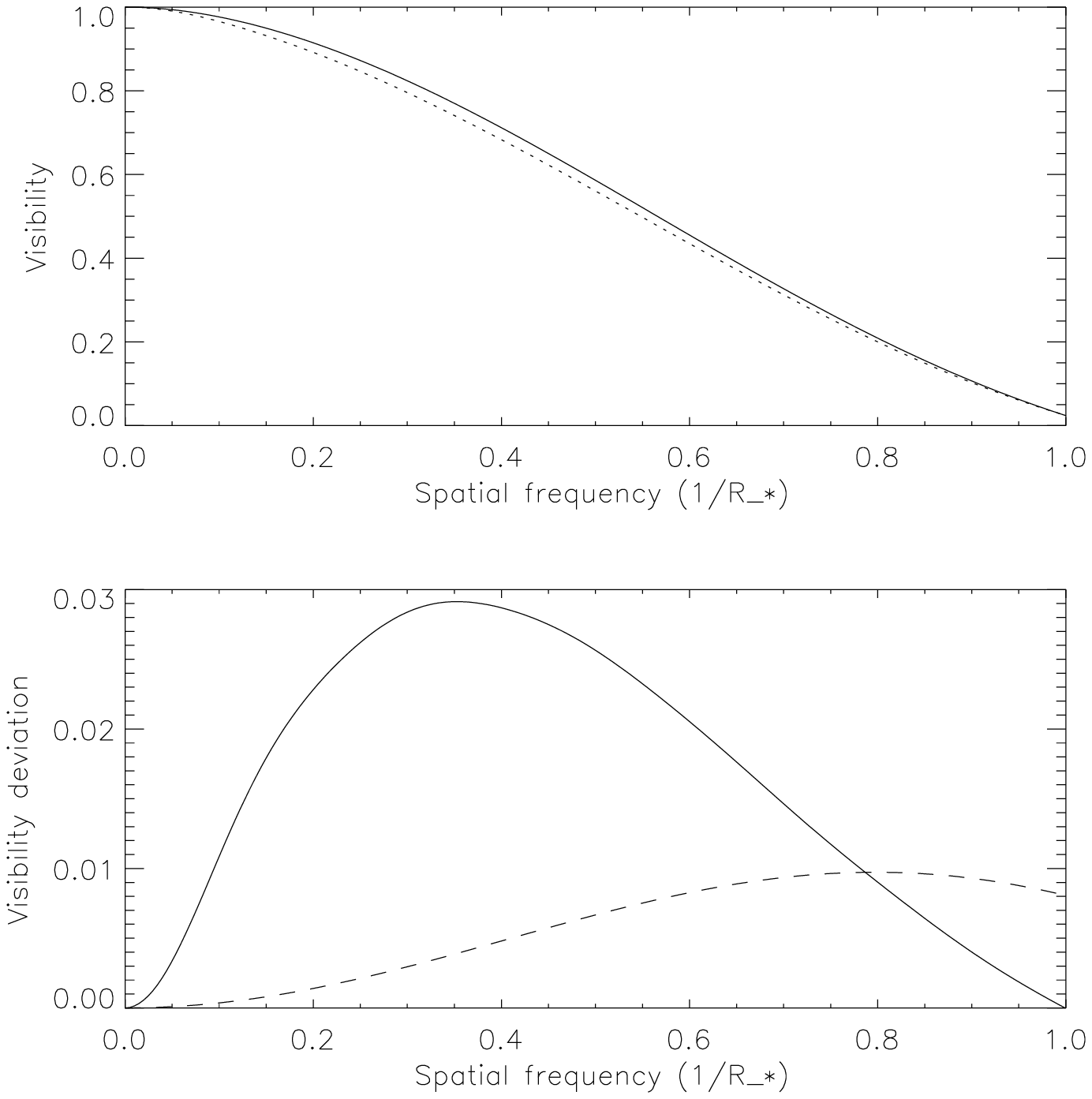}
   \includegraphics[height=5.3cm]{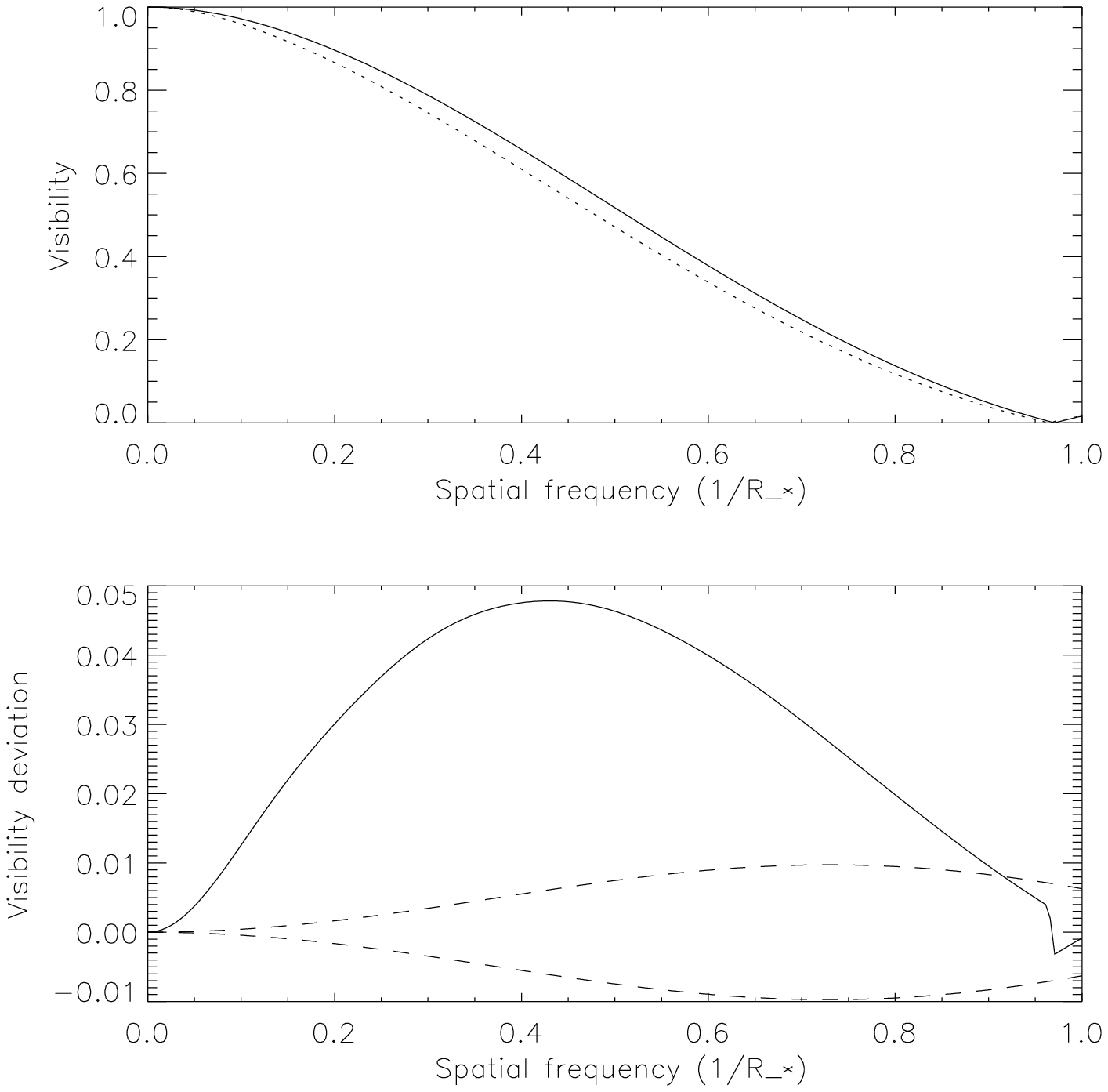}
   \includegraphics[height=5.3cm]{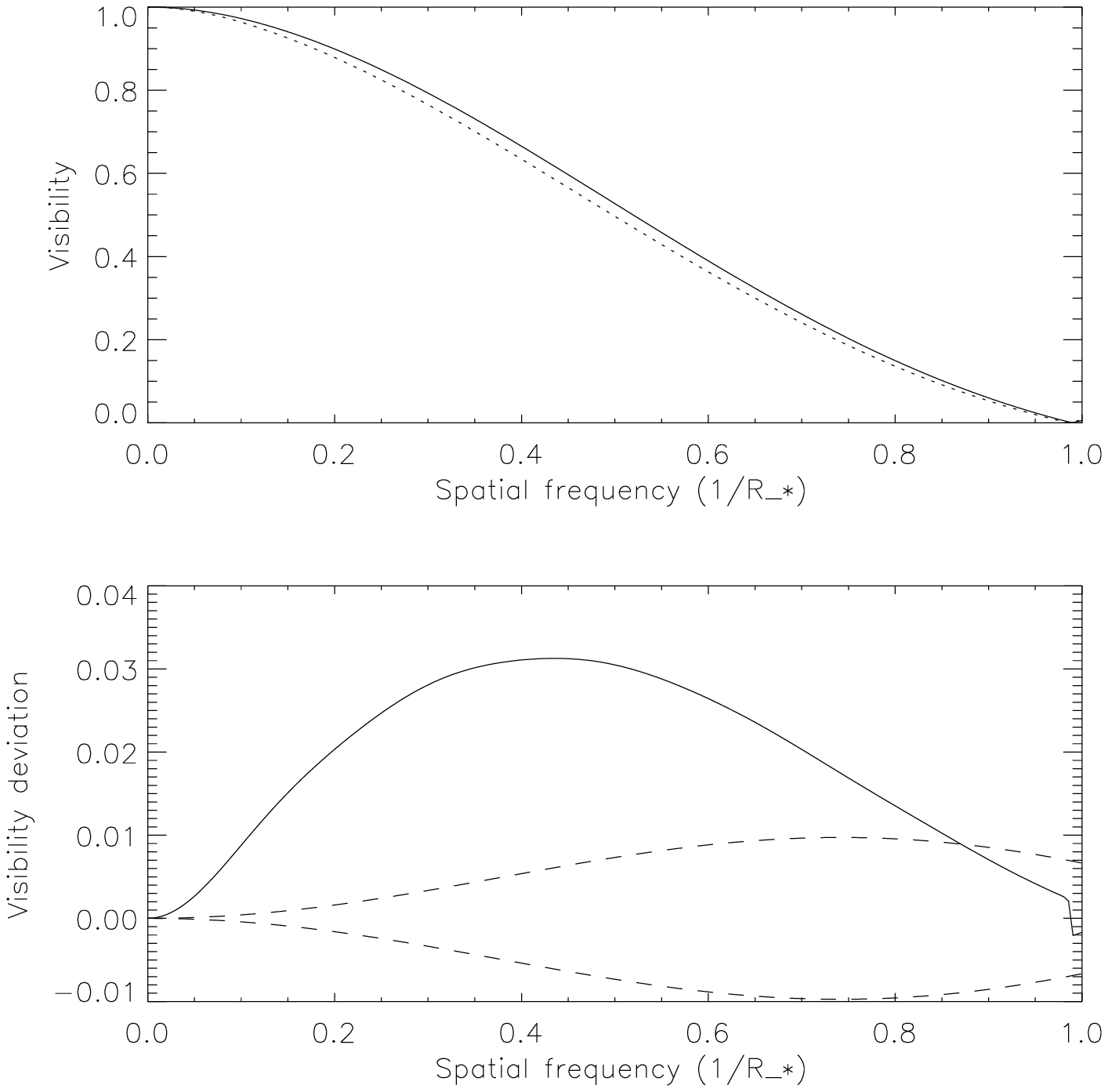}
   \caption[]
  {\label{fig:ZT}
Model of $\zeta$ Tau for 3 inclinations corresponding to 3 disk
inclinations (columns). The upper subpanels show the intensity
maps on the sky with the overplotted polarization $P$. The thick
bar represents a linear polarization of 100\%. The lower panels
present the overplotted visibility curves for $I_{\rm s}$ and
$I_{\rm p}$ maps ($s$ polarization in solid line and $p$
polarization in dotted line), and their difference $\Delta V_{\rm
P}$ for a vertical ($z$, middle panels) and horizontal ($y$,
bottom panels) baseline orientation. For each of the $\Delta
V_{\rm P}$ curves corresponds a couple of 1\% sensitivity curves
(dashed) defined in Eq.~\ref{eq:sens} to illustrate the strength
of the signal.}
 \end{figure*}

\subsection{Be stars}
\begin{figure*}
  \begin{center}
      \includegraphics[height=6.2cm, width=16.0cm]{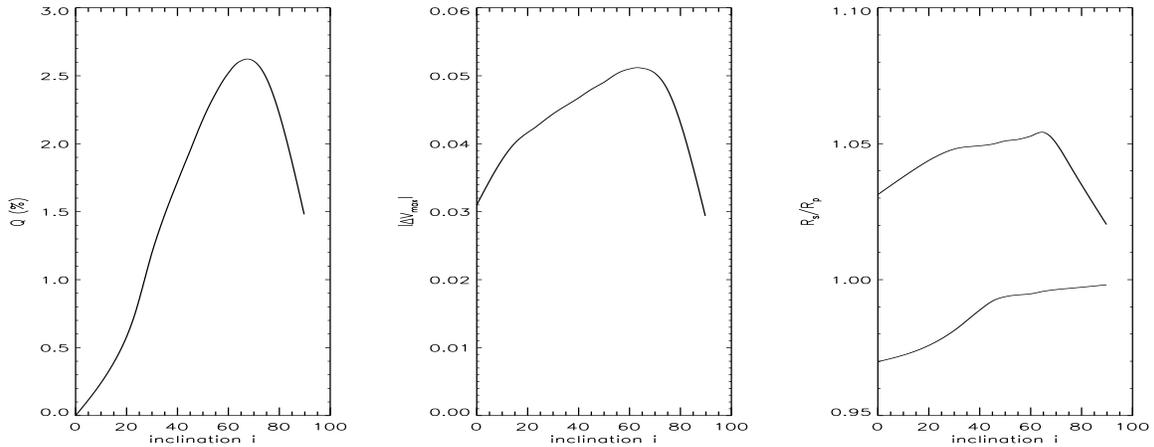}
  \end{center}
  \caption[]{\label{fig:SPIN}
    {\sl Left:} Polarimetric and SPIN signal for various inclinations.
    {\sl Middle:} Expected SPIN signal, represented by the maximum of
    the deviation curve (as shown in Fig.\ref{fig:ZP-standard2}).
    {\sl Right:} Ratio of uniform diameters (UD) fit between two perpendicular polarization
    direction. The upper (lower) curve describes the
    ratio for an horizontal (vertical) baseline.
    Due to the spherical symmetry, the pole-on signal
    is just inverted between the baseline. At higher inclinations $r_{\rm H}$ increases, but
    the vertical polarized signal disappears and $r_{\rm V}$ reaches almost 1.}
\end{figure*}

Be stars are hot and fast rotating stars surrounded by an extended
circumstellar hydrogen envelope. They manifest the so-called 'Be
Phenomenon' characterized by Balmer lines in emission and infrared
excess. One of the challenging questions on Be stars is the
geometry of their disk, and in particular their opening angle,
about which there is still an active debate. Most authors have
considered geometrically thin disks (half opening angle of
2-5$^{\rm o}$). The very narrow disks considered by Wood et al.
(1997) were those predicted by the
 Wind-Compressed Disks (WCD) theory of Bjorkman and Cassinelli (1993).
Furthermore, interferometric observations have given upper limits
of approximately 20$^{\rm o}$ ~\cite{Quirrenbach97}. However, the
hypothesis of such narrow disks faces several problems, and the
current set of observations
does not provide a unique interpretation on the circumstellar geometry (Yudin et al. 1998).\\
The model from Waters~\cite{Waters86} has been successfully used
to explain the near and far IR observations and is coherent also
with polarization data (Cot\'{e} \& Waters 1996, Waters \&
Marlborough 1992). They model the disk as an equatorial cone with
a density law described in Eq.~\ref{eq:power} with a density
gradient $n \sim 2-3.5$, the density of the disk $\rho_0 \sim
10^{11}-10^{13}$~cm$^{-3}$, the disk radius $R_{\rm d}$, the
viewing angle $i$, and the disk half-aperture $\theta$ as main
parameters. We performed simulations of the polarized emission of
Be star disks. These simulations have been tested using the work
from Wood et al. (1996, Fig.~\ref{fig:SPIN}). We obtain very
similar results for the integrated polarization, (with less than
10\% deviation) taken into account that our density law is not
exactly similar to that used by Wood et al. which allows a
colatitude dependency of the density. However, their
polar-to-equatorial density ratio of $1:10^3$ is very strong, so
that we can consider the two models being very close. In
particular, we used as template the star $\zeta$ Tau (B1 IVe-sh
star), which is a well studied case of an edge-on Be star
($i\simeq82^\circ$; Wood et al. 1997). We
 used a set of parameters close to those of Wood et al.~\cite{Wood97}, i.e.,
$R_* = 6 R_\odot$, $T_* = 2\times10^5~{\rm K}$ for the central
star and a distance of 120pc. For the disk simulation a very thin
disk of half-opening angle of 3$^{\rm o}$ and density law which
defines an optical depth in the disk plane of about $\tau_{\rm
e}=3$ (density gradient $n=-3$) are used. We recall that in our
study the only emission process considered is Thomson diffusion.

In Fig.~\ref{fig:ZT} we present the expected interferometric
signal for $\zeta$ Tau for different inclinations. In the pole-on
case, we clearly see that the visibility deviation curves $\Delta
V_{\rm P}(f)$ are identical but inverted between both baseline
(the baseline is aligned and perpendicular to the polarizer). This
configuration is the most favorable for the local polarization:
the bulk of electrons is perpendicular to the line of sight and
$\Delta V_{\rm max}$ reaches 0.03. For $i=45^\circ$, the local
polarization in the vertical direction is much weaker. This is due
to the thinness of the disk: for most of the electrons, the
polarization efficiency is only 30$\%$ at the diffusion angle
$\chi=$45$^\circ$ ($P=(1-\cos(\chi)^2) / (1+\cos(\chi)^2)$). The
SPIN signal is almost undetectable but for the perpendicular
baseline, the signal is much stronger ($\Delta V_{\rm
max}=0.045$). We see in this example that the ratio of $\Delta
V_{\rm P}$ between two perpendicular baselines provides valuable
information on the system inclination $i$ on the sky and about the
aperture of the disk. This complements the natural light
information, i.e.\ the ratio of the radii in two perpendicular
directions which provides information on the projected 2D
intensity
map but no indication on the real 3D structure of the object.\\
In the equator-on configuration, an unexpected large signal
($|\Delta V_{\rm max}|=0.02$) is visible at a high spatial
frequency ($f_{\rm max}=0.83)$ with a vertical baseline (i.e.
perpendicular to the disk). This effect is related to the
increasing disk vertical extension in the external regions and is
strengthened by the absorption of the unpolarized star light in
the line of sight. With an horizontal baseline, $f_{\rm
max}=0.37$, i.e.\ lower than in the previous case, and $\Delta
V_{\rm max}=0.03$. This is also due to the optical thickness of
the disk, which prevents the observation of high polarization
close to the line-of-sight to the star. The bulk of polarization
is thus located further out, and $f_{\rm max}$ is decreased
compared to other inclinations.

Fig.~\ref{fig:SPIN} illustrates the differences and the
complementarity between the polarimetric and interferometric
observables. We reproduce quantitatively the integrated
polarization curve from Wood et al.~\cite{Wood96}. In edge-on
view, there is a strong decrease of the local polarization due
  to the highly optically thick line of sight (multiple scattering).
  Nevertheless the integrated polarization is compensated by the
  higher asymmetry of the system (right panel in the figure).
In pole-on view the integrated polarization is null but the SPIN
signal is still detectable and even slightly larger than in the
equator-on case. For integrated polarization and SPIN signals the
maximum occurs at the same angle $i\simeq65^\circ$. The
interferometer is very sensitive to the large projected emitting
surface of the pole-on view which compensates the lower local
polarization.

\subsection{B{[}e{]} stars}
B{[}e{]} stars are hot supergiants showing an important excess in
the infrared due to the presence of hot circumstellar dust. These
stars exhibit also the so-called 'Be Phenomenon', but also show
forbidden lines in their spectrum. Zickgraf et al. (1985) proposed
a model for the LMC B{[}e{]} supergiant R126 consisting of a fast
wind in the polar regions and a dense and slow wind in the
equatorial region where the dust is formed. In contrast to Be
stars, the circumstellar geometry of B[e] stars is rather more of
an open question. Moreover, within the sample of classified
galactic B{[}e{]} stars which could be resolved by an interferometer,
one can find young stellar objects such as extreme Herbig Be stars
(HaebeB{[}e{]}), together with supergiant stars (sgB{[}e{]}, see
Lamers et al. 1998). The distance estimations and hence their
luminosity and angular diameter are poorly constrained, which
render the classification problem.
\begin{figure}
  \begin{center}
      \includegraphics[height=8.cm]{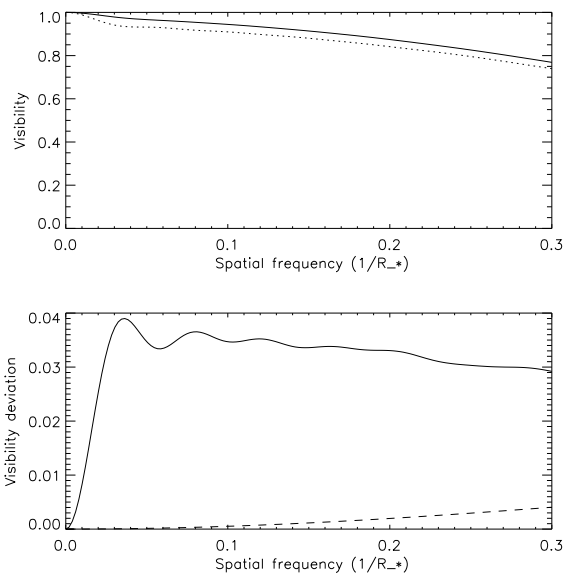}
  \end{center}
  \caption[]{\label{fig:BeC}
Model for a B{[}e{]} star environment without dust: $N_{\rm
e}=6\times10^{9}$
 cm$^{-3}$ (constant density), $\theta=10^\circ$, $i=90^\circ$, $R_*=70{\rm R}_\odot$,
 $R_{\rm outer}=40{\rm R}_*$, $d=2$kpc. $\tau_{\rm e}=0.8$ in the disk plane, and the system is seen
 edge-on. We show a close-up of the two visibility curves $|V_{\rm s}|$ (solid line) and
 $|V_{\rm p}|$ (dotted line). The baseline is aligned to the disk. The contribution of polarized light
from the envelope is clearly visible in the oscillation of the
polarized differential visibility curve $\Delta V_{\rm P}$ and the
signal is well above the 1\% sensitivity of a state-of-art
interferometer illustrated by the dashed curve. The diffused light
represents 8\% of the total flux and is almost completely linearly
polarized which explains why the envelope signal is concentrated
only in one polarization.}
\end{figure}
As revealed from their emission lines and near IR (J band) excess,
the envelopes of sgB{[}e{]} stars provide an ample opportunity for
scattering of radiation from the central star by free electrons.
Moreover, dust is evidenced by their IR excess (Melgarejo et al.
2001). It is possible but sometimes difficult to discriminate the
electron scattering and the dust scattering regions with
spectropolarimetry alone (Oudmaijer \& Drew 1999) and our SPIN
technique offers a great help to disentangle between different
extents of the polarizing sources. It is not on the scope of this
paper to perform a detailed and complex modelling of several
examples of sgB{[}e{]} stars. We show here the expected impact on
the SPIN signal of a typical B{[}e{]} stars environment, i.e.\ an
extended, low density scattering region, optically thin and
somewhat flattened by using a toy model adapted from Melgarejo et
al. 2001.
\\
For that purpose, we use the same model as for the Be stars with a
diluted environment (typically $N_{\rm e}\sim 10^{9}$ cm$^{-3}$),
more open($\theta \sim 5^\circ-20^\circ$) and extended (outer
radius $R_{\rm outer} \sim 40R_* - 300 R_*$), without any dust.
The result for an edge-on inclination is shown in
Fig.~\ref{fig:BeC}. The principal characteristic of this
environment is the angular diameter contrast between two distinct
flux sources: the star and a faint but highly polarized extended
envelope. The envelope contribution can be easily seen in
Fig~\ref{fig:BeC}. When the baseline and the polarizer are
oriented both in the disk direction, the visibility curve is
completely dominated by the stellar flux. When the polarizer is
aligned with the $p$ direction, the bulk of the envelope becomes
visible, superimposed on the stellar component. The envelope is
highly resolved so that its contribution is detectable at a low
spatial frequency: $\Delta V_{\rm max}=0.04$ at $f_{\rm
max}=0.036$ for the model presented in Fig.~\ref{fig:BeC}. The
integrated polarization $P$ amounts to 1.8\%. The signal amplitude
is close to one of the Be star, but its shape is very different
due to the contrast between the point-like central source and the
very extended and diluted environment.
\\
In conclusion, for B{[}e{]}, we expect that the SPIN signal from a
polarized (by electrons or dust scattering) detached envelope will
be easier to extract from the stellar component due to the
contrast between star and envelope extents. The envelope polarized
and natural relative contribution, its extent and geometry could
therefore be retrieved from a relatively simple model of its
extracted visibility curve. For Be stars such a simple reduction
process is complicated by the fact that the disk extent is much
lower, and because the disk is supposed to be highly optically
thick for most of the models with small aperture.


\section{Instrumental Application}
\label{sect:ins} After a long development, optical interferometry
is now ready to play a significant role in astronomy. The
emergence of well-funded interferometer 'facilities ' allows to
enlarge the field of applications of this technique considerably,
by increasing its reliability and its sensitivity. Observing in
polarized light with interferometers promises fascinating new
insights into many areas of astrophysics, although this capability
is difficult to implement with current interferometers. The
instrumental polarization in interferometers has been studied by
Rousselet-Perraut (1996) and Elias (2001), and we suppose in this
section that the SPIN instrument can control and calibrate the
effect of the internal polarization on the SPIN observables.

In principle there is no obstacle to equip the focal instrument of
an interferometer by polarimetric optics. A simple polarization
analyzer using a Wollaston prism can perfectly match the
specifications for calibration and visibility determination in the
different polarizations. In practice we record fringe patterns in
linear polarizations and estimate complex visibilities for each of
them. The most promising way to calibrate the signal is to use a
differential technique (i.e. by cross-correlating the signal from
both polarization directions for instance). The polarized
deviation curve can therefore be determined with greater accuracy.
It is necessary to use an unresolved star as reference to estimate
the absolute visibility. This step can be perfectly carried out
only in natural light if the polarized signal can be considered as
second-order effect. The first observations have to use large
spectral bandwidth to optimize the sensitivity for which a 1\%
accuracy is expected routinely in natural light.

In the current state-of-the-art of interferometers, it is time
consuming to record many visibility points with different baseline
lengthes and directions. For spherical targets, the visibility in
natural and polarized light does not depend on the projected
baseline direction, but only on its length. It means that the data
recorded with similar baseline lengths can be added in order to
increase the SPIN SNR, since the polarizer direction of analysis
follows the baseline movement during observation.  For Be and B[e]
stars, the measured visibility (in natural or polarized light)
does in general not longer depend on the direction of the
projected baseline (except for pole-on configurations). The
visibility changes with the baseline movement (earth rotation)
which restricts the number of visibility points recordable per
independent configuration compared to the spherical case. This
problem is compensated by the large signal expected from these
stars as seen in Sect.\ref{sec:2D}. It can be somewhat difficult
to overcome the degeneracy between disk density, aperture and
inclination but the polarized visibility provides an complementary
information useful to constrain the parameter space.

\begin{table*}[]
\caption[]{Comparison of the SPIN signal between two wavelengths.
Be stars radii are estimated from Quirrenbach et al. 1997, and
P~Cygni radius from this study. The wavelength correction factors
are estimated from spectropolarimetric observations: Quirrenbach
et al. 1997 for Be stars, and Nordsieck et al. 2001 for P~Cygni.
In the $f_{\rm max}$ columns the corresponding baselength are
indicated in units of meters. In the $\Delta V_{\rm 50}$ columns
the expected $\Delta V_{\rm P}$ for a 50\,m baseline are
reported.}\label{ta:be_wave}

\begin{tabular}{|l|c|c|c|c|c|c|c|c|}
\hline Name & Distance & $\Theta_{\rm ap}$  &
\multicolumn{3}{|c|}{$\lambda_\mathrm{eff}$=0.55$\mu {\rm m}$} &
\multicolumn{3}{|c|}{$\lambda_\mathrm{eff}$=0.66$\mu {\rm m}$} \\
    & in pc & in mas & $\Delta V_{\rm max}$&
$f_{\rm max}$ &$\Delta V_{\rm 50}$&
$\Delta V_{\rm max}$& $f_{\rm max}$ & $\Delta V_{\rm 50}$\\
\hline
$\gamma$ Cas & 190 & 0.56 &0.05-0.06& 100 &$\simeq$0.02&0.03-0.04& 120 & $\simeq$0.015\\
$\zeta$ Tau & 130 & 0.4 & 0.02-0.03 & 130&$\simeq$0.015&0.015-0.025 & 160& $<$0.01\\
$\eta$ Tau & 110 & 0.71 & 0.03-0.04 & 40 & $\simeq$0.035 &0.025-0.03& 50 & $\simeq$0.025\\
P Cygni & 1800 & 0.55 & 0.08-0.09 & 100 &$\simeq$0.05&0.065-0.075 & 120 & $\simeq$0.04\\
\hline
\end{tabular}
\end{table*}

Which is the optimum wavelength region for this study? From the
polarimetric point of view, the polarized flux generated by
Thomson scattering decreases generally in the IR domain due to the
competing influence from free-free continuum optical depth
$\tau_{\rm ff}$. Since $\kappa_{\rm ff} \propto a_{\rm ff}
\lambda^2$, the local polarization decreases as $e^{-a_{\rm ff}
\lambda^2}$. For instance, the integrated polarization of P~Cygni
is decreased by a factor 3 between 0.55 $\mu {\rm m}$ and 1 $\mu
{\rm m}$ (Nordsieck et al. 2001). It is therefore more interesting
to observe towards shorter wavelengths. For Be stars the
bound-free opacities cannot be neglected. In Tab.\ref{ta:be_wave},
we present semi-quantitative signal expectations for Be stars and
P~Cygni computed by means of the spectro-polarimetric data
available for these stars. As an example, the spatially integrated
polarization of $\gamma$ Cas declines from 0.6\% at the Balmer
jump, to reach 0.52\% at 0.5$\mu{\rm m}$, and 0.4\% at 0.66
$\mu{\rm m}$. This evolution reflects only the changes in the
free-bound opacity and free-free emission towards the envelope and
affects the local polarization and the SPIN signal. The
differences between the spectrophotometry of $\eta$ Tau (nearly
pole-one), $\gamma$ Cas (i$\simeq 45^\circ$) and $\zeta$ Tau
(nearly edge-on) are mainly due to inclination effects.

The IR domain is also less attractive in term of spatial
resolution. This is particularly striking in the context of hot
stars since even for the examples presented in this paper, the
minimum baseline range needed to resolve the polarized environment
is 150-200m, i.e.\ at the upper limit of possible VLTI baselines. The number of
available baselines for such a scientific task is thus
dramatically decreased.

On the other hand, the disturbing effect from the atmosphere is
more striking in the optical, and the interferometers able to
perform observations in the optical wavelength range are currently
few, and somewhat less sensitive than NIR ones. We nevertheless
estimate that the gain in spatial resolution and polarized signal
is such that the visible wavelength range should be preferred for SPIN observations
of hot stars environments, at least for electron scattering
studies. We also want to mention the numerous applications of
SPIN for the study of dusty environments studies, which are not
directly in the scope of this paper. For these environments, the
requirements in terms of spatial resolution are strongly
decreased, and the IR domain is best suited.

At that moment, only one long-baseline interferometer, the GI2T,
is equipped with a polarimeter device for routine observations.
The GI2T-REGAIN spectro-interferometer is composed of two 1.5-m
telescopes which can be displaced on a North-South baseline
spanning from 12m to 65m. The REGAIN beam combiner forms the focus
in visible light and is equipped with a visible spectrograph in
whch a polarimetric device can be inserted. For a complete optical
scheme and the status of the instrument, see Mourard et al. 2000a
and Mourard et al. 2002. The technical description of the
polarimetric observation can be found in Rousselet-Perraut et al.
2002.

 In the near future, further instruments will provide polarimetric
facilities. For instance, the foreseen near-infrared AMBER/VLTI
could observe, in a first phase, with a polarimetric device which
processes only half of the incoming light. This device has been
implemented in order to control the instrumental polarization, but
this technical constrain could become a great opportunity to test
the SPIN technique since it allows to extract directly the $\Delta
V_{\rm P}$ parameter.

A concept of a polarimetric interferometer for the Very Large
Telescope Interferometer (VLTI), VISPER (VLTI Imaging
Spectro-PolarimetER) has been presented by Vakili et al. 2001.
Since great care was taken on instrumental polarization effects
during the design and construction of the VLTI, polarimetry could
be straightforwardly implemented in the interferometric
laboratory. The large baselines, high-order adaptive optics (in
the optical), fringe tracking and the foreseen dual-field facility
PRIMA can greatly enhance the sensitivity of an
interfero-polarimeter to the measurement of small polarization
effects on the visibility and extend the number of stars for which
a SPIN signal can be detected. Finally, the SPIN technique implies
that the observations are carried out with a certain spectral
resolving power. This resolving power could allow a study of the
visibility through stellar spectral lines in a similar way as
spectropolarimetry. The differential information provided by a
comparison between the continuum and the lines is richer and more
sensitive than the one provided by classical interferometric
technics. It allows to retrieve the differential phase
information, related to the evolution of the stellar photocenter
through the spectrum, and also a dynamic information on the
polarized environment through a spectral line (see Vakili et al.
1997, Berio et al. 1999). With PRIMA the calibration can be even
carried out in an absolute way. As an example, the hot star
emission lines can often be considered as unpolarized and offer a
good opportunity to differentially calibrate the polarized
continuum, especially for WR stars. The dynamically complex
environment of Be stars could also be studied with SPIN. Poeckert
\& Marlborough (1978) have demonstrated that the H$\alpha$ line is
polarized and contains information on the disk structure and
dynamics. This subject should be investigated
in a future work.\\



\section{Conclusions}
Spectro-Polarimetric INterferometry is a new and complementary way
to study the polarized light from hot star environments. The
purpose of this article is to present an overview of the potential
of the SPIN technique, limited to the study of Thomson scattering
within the wind of hot stars. The main point of the SPIN concept
is that the star appears to have a different radius depending on
the orientation of the polarizers relative to the baseline. This
investigation demonstrates that state-of-the-art optical
interferometers can detect the visibility difference (the 'SPIN
signal') for a broad range of stellar wind parameters. The signal
is particularly large for the denser winds and can be detected
with small baselines (50-150m). For stars exhibiting dense winds,
the envelope is extended and polarized by a large amount whereas
the direct, mostly unpolarized stellar light is damped by the
scattering, and only barely resolved by the interferometer. The
SPIN technique also provides a great wealth of information on Be
star disks: disk radii in natural light for different baselines,
the difference of visibilities between two direction of
polarization {\it and} baseline orientations. When the spatial
scales of the polarized environment and the central star are
different, the extraction of the spatial information is
simplified, as seen for B{[}e{]} stars.

This work has also to be seen in a larger context. The signal
modelled in this study is certainly not the largest expected from
non-extended astrophysical targets. The environments described
here are geometrically and physically similar to astronomical
objects ranging from the dusty environment of Young Stellar
Objects, AGB stars to the complex environment of AGNs whose
apparent angular diameter is equivalent to the targets presented
here. Thus, we strongly advocate the development of
interferometric devices dedicated to SPIN measurements within the
frame of second generation VLTI instruments.

\begin{acknowledgements}
The authors thank J.P. Aufdenberg for having provided a numerical
detailed model of Deneb. O.C.\ is grateful to the ``Max-Planck
Institut f\"ur Astronomie`` for a postdoctoral grant. S.W.\
acknowledges financial supported through the NASA grant
NAG5-11645, through the grant DP.10633 (project 101555) and
through the SIRTF Legacy Science program through an award issued
by JPL/CIT under NASA contract 1407. A.D.S. acknowledges CAPES -
Brazil (contract BEX 1661/98-1) for financial support. O.C.
dedicates this work to his new born son Mathieu and his wife
Martine.

\end{acknowledgements}

\end{document}